\documentclass[%
  superscriptaddress,
  twocolumn,
  10pt,
  aps,
  pre,
  nofootinbib, 
  floatfix,
  showkeys
]{revtex4-2}

\usepackage{hyperref}
\usepackage{uri}
\usepackage{graphicx}
\usepackage{amsmath, amssymb}
\usepackage{xfrac}
\usepackage{multirow}
\usepackage{makecell}
\usepackage{todonotes}
\usepackage{printlen}
\usepackage{xcolor}

\DeclareMathOperator*{\argmax}{argmax}

\begin{abstract}
Unknown node attributes in complex networks may introduce community structures that are important to distinguish from those driven by known attributes. We propose a \emph{block-corrected} modularity that discounts given block structures present in the network to reveal communities masked by them. We show analytically how the proposed modularity finds the community structure driven by an unknown attribute in a simple network model. Further, we observe that the block-corrected modularity finds the underlying community structure on a number of simple synthetic network models while methods using different null models fail. We develop an efficient spectral method as well as two Louvain-inspired fine-tuning algorithms to maximize the proposed modularity and demonstrate their performance on several synthetic network models. Finally, we assess our methodology on various real-world citation networks built using the OpenAlex data by correcting for the temporal citation patterns.
\end{abstract}

\begin{document}
\title{Block-corrected Modularity for Community Detection}
\author{Hasti Narimanzadeh}
\email{hasti.narimanzadeh@aalto.fi}
\author{Takayuki Hiraoka}
\author{Mikko Kivelä}
\affiliation{Department of Computer Science, Aalto University, 00076 Espoo, Finland}


\maketitle

\section{Introduction}
Complex systems typically consists of elements interacting with one other, commonly represented by networks of nodes connected by edges. Many such systems comprise multiple types of elements that differ in their characteristics, affiliations, roles, or functions. For example, proteins in cells accomplish different biological functions; species in an ecosystem occupy different niches and habitats; individuals in society can be categorized by gender, age, educational background, or political beliefs; scientific articles vary in topics, fields of study and institutional affiliations of their authors.
These node-level \emph{attributes} can be known, i.e., directly observable, or unknown, i.e., some characteristics of the nodes which are not directly measured.
The likelihood that nodes are connected is often influenced, to different extents, by their attributes through mechanisms such as homophily or assortativity. That is, two elements that share functions or characteristics are more likely to interact. When mechanisms are present that translate attributes into network structure, those attributes can be inferred from the network topology even when they are not directly observable---a task commonly referred to as community detection. 
Indeed, the communities are typically interpreted as proxies for some unknown attributes, such as functional roles of genes or social groups of individuals~\cite{girvan2002community, ji2012survey, traud2012social, guerra2013measure, hric2018stochastic, salloum2022separating}.

Among the many community detection methods that aim to uncover unknown attributes, many of the most popular approaches are based on \emph{modularity}~\cite{newman2006finding}. Built upon maximizing the modularity quality measure across all possible partitions, modularity optimization methods comprise a vast collection of algorithms, ranging from greedy algorithms~\cite{traag2009community, blondel2008fast} to spectral methods~\cite{newman2006finding, richardson2009spectral, white2005spectral}.
Modularity, in essence, measures the abundance of edges within each community as compared to that predicted by a randomized \emph{null model} of the network with no community structure. Despite its prevalence, modularity maximization is not without its shortcomings and has faced criticism over the recent years~\cite{peixoto2021descriptive, good2010performance}. Nonetheless, it is a versatile method easily adaptable to various network types such as multilayer, temporal multilayer, spatial, and motif networks to name a few~\cite{speidel2015community, de2013mathematical, bazzi2016community, expert2011uncovering, liu2012extending, arenas2008motif}.

Detecting communities in networks whose every node is annotated by metadata comprises another vast arena. These metadata are observable (known) node attributes that can either be continuous, discrete or categorical. On many occasions, it has been shown that such node metadata can be correlated with the connectivity patterns of networks~\cite{bothorel2015clustering, chunaev2020community, red2011comparing, traud2012social}. In light of these observations, a widespread assumption emerged that these known attributes intrinsically align with the unknown attributes, which has led to the development of methods that ensure homogeneity of known attributes in the community structures that have yet to be uncovered~\cite{jia2017node, asim2017community, zhou2009graph, huang2017joint}. A common approach, for example, is to construct a similarity network based on the known node attributes, then perform community detection, e.g., using modularity-based methods, on its union with the original network~\cite{jia2017node, zhe2019community, asim2017community}.

It is not, however, always the case that known attributes are the sole drivers of the network structure, or that the unknown attributes that are not directly observed in data align with the known ones. If such unknown attributes induce a structure that is distinct from that driven by the known ones, focusing on the latter may eclipse the former. That is, conventional community detection methods, in such scenarios, would either return the known attributes or yield communities that are distorted versions of the unknown attributes. Thus excluding the influence of known attributes on the structure and retrieving the community structure underpinned by unknown attributes will lead to more meaningful findings.

\begin{figure}[!tb]
  \centering
  \includegraphics[width=0.89\linewidth]{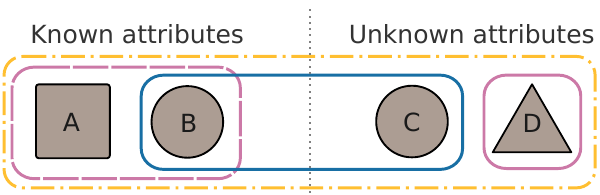}
  \caption{A schematic of the relation of community structures found by three types of community detection methods to the attributes that induce assortative structures in a network. In this example, nodes are labeled with two known attributes (A and B) and two unknown attributes (C and D) where attributes B and C are closely aligned.
  Using a conventional community detection method with no consideration for metadata, the discovered structure is affected by all known and unknown attributes A, B, C, and D (dot-dashed, yellow in color). If a method assumed a specific known attribute, in this example B, was aligned with the community structure, the found communities would resemble the effect of this known attribute and the unknown attribute C with which it is aligned (Dark solid, blue in color). Our proposed method would nullify the effect of the known attributes A and B (dashed, purple in color) as well as the unknown attribute C to arrive at a community structure solely driven by the unknown attribute D (light solid, purple in color).}
  \label{fig:schematic-attributes}
\end{figure}

Academic citation networks provide an example of how known attributes can obscure potentially intriguing unknown ones. The publication time of each article imposes a rather strong constraint on its citation behavior, as papers generally cannot cite future publications, nor do they cite too recent or too old papers~\cite{parolo2015attention}.
Therefore, what we discover by applying off-the-shelf community detection methods may strongly reflect the temporal connectivity of the network, as opposed to other driving forces that may be at play, such as academic disciplines. At times we may be able to use the known attributes to bias the discovered communities towards what we are interested in---for example, the journal in which an article is published can be treated as node metadata that aligns with the academic field. This approach, however, relies heavily on the assumption that there is an available known attribute that aligns with the unknown one, which is not necessarily the case. If publication date and journal are the only known attributes, previous methods cannot uncover the community structure caused by other attributes, such as the countries where authors are based. We illustrate this schematically in Fig.~\ref{fig:schematic-attributes}.

Here, we take the opposite approach: for a given network data accompanied by node metadata, we discount the effect of known attributes on network structure in order to isolate the effect of unknown attributes. This amounts to asking whether there exists any community structure beyond the known attributes. 
Factoring out the known attributes has been previously accomplished in spatial networks, for example, by adapting the definition of modularity to discount the effects of spatial embedding~\cite{expert2011uncovering}. In a mobile phone network where calls are predominantly placed among persons in close physical proximity, nullifying this physical proximity of callers came to reveal linguistically homogeneous communities, communities of callers speaking the same language. Similar approaches have been proposed for temporal networks by accounting for an ordering on the nodes, in either modularity-based~\cite{speidel2015community} or Bayesian inference methods~\cite{peixoto2022ordered}.

In this work, we propose a modularity measure with a stochastic block model as a null model that can factor out any categorical attributes. We illustrate that the model isolates and captures the effects of unknown attributes on the network structure by incorporating arbitrary connectivity patterns induced by known attributes. We also derive a fast, vectorised and efficient extension of the spectral modularity maximization heuristic~\cite{newman2006finding}, as well as two greedy fine-tuning algorithms.
We demonstrate the utility and scalability of our method on real-world citation networks containing millions of nodes, with publication time as the known attribute, showing that null models proposed in prior studies often impose specific assumptions of connectivity which might not align with the observed empirical citation patterns, especially considering how they have evolved in more recent times. In contrast, our proposed method offers the flexibility to accommodate arbitrary connectivity patterns, providing a more adaptable framework for community detection.

First, in Sec.~\ref{sec:block-corrected-definition} we formally define our proposed null model and modularity. Next, in Sec.~\ref{sec:examples} we show with two sets of examples how this null model can capture the unknown community structure even in the presence of known attributes that significantly affect the connectivity of the network. Next, in Sec.~\ref{sec:heuristics} we discuss the practical application of our null model on large networks, specifically the intricacies of efficiently applying our proposed null model to sparse networks. Finally, in Sec.~\ref{sec:openalex} we demonstrate the utility and the efficacy of our proposed null model on real-world citation networks ranging from hundreds of thousands to tens of millions of edges.

\section{Block-corrected modularity}\label{sec:block-corrected-definition}
In the following, we focus on unweighted, directed networks characterized by their asymmetric adjacency matrix $\mathbf{A}$. Element $A_{ij}$ of the adjacency matrix $\mathbf{A}$ equals one if there is an edge from node $i$ to node $j$ and zero otherwise. $V$ is the set of nodes in the network. The number of edges and nodes in the network is denoted by $m$ and $|V|$, respectively.

Behind most community detection methods lies the fundamental idea that communities of nodes in the network are densely connected within and sparsely connected between communities. Associated with this notion is often a mathematical definition that assesses the quality of a candidate partition of the network. These are associated with optimization methods aiming to maximize the quality function across all possible node partitions. One of the most widely recognized quality functions is the modularity function \cite{newman2006finding} defined mathematically as
\begin{equation}\label{eq:modularity_func}
    Q = \frac{1}{m} \sum_{C \in \mathcal{P}} \sum_{i, j \in C} \left(A_{ij} - P_{ij} \right)\,,
\end{equation}
where the summation $i, j \in C$ is over all pairs of nodes $i$ and $j$ in a community $C$ in a candidate partition of the network denoted by $\mathcal{P}$. This summation applied to the corresponding elements of the adjacency matrix $\mathbf{A}$ equates the number of edges in community $C$. Modularity value $Q$ tells how salient the communities in the candidate partition are compared to a random reference network. The notion of a random reference network is captured by the parameter $P_{ij}$, which is the expected probability of an edge from node $i$ to $j$ across an ensemble of random networks, a generative \emph{null model} with certain constraints assimilated into edge probability $P_{ij}$. These constraints embody specific features of the network under study which must be preserved in the null model, such that generated random networks are maximally similar to the original network without containing any community structure other than what is induced by the known attributes.

Multitudes of choices for null models have been developed over the years, ranging from Erdős-Rényi $G(n, p)$ model where number of nodes and the edge probability are constants, preserving the expected number of edges~\cite{erdds1959random, erdHos1960evolution}, the configuration models~\cite{molloy1995critical, bollobas1980probabilistic, holland1983stochastic} wherein degrees of nodes are preserved, to models designed for bipartite networks~\cite{Barber2007ModularityCommunity}, spatial networks~\cite{expert2011uncovering}, directed acyclic networks~\cite{karrer2009random, speidel2015community}, models designed for uncovering communities in networks with signed edges~\cite{traag2009community, pougue2024signedlouvain} or generating dynamic networks with temporal motifs~\cite{zeno2021dymond, purohit2018temporal}. 
All such models are characterized by preserving core qualities present in the observed network in their respective null models, which is integral to uncovering unknown community structures as we will expound shortly.

\subsection{Block null model}
Widely used as an inference framework, stochastic block models~\cite{karrer2011stochastic, holland1983stochastic} also serve as effective generative models, much like a null model is. They are characterized by their block connectivity matrix which can be interpreted as an array of discrete node-level attributes. In the study of real-world networks, it is often the case that some attributes are known \emph{a priori}, while some are unknown. Unknown attributes may impose a community structure that can be of interest to retrieve and may be largely independent of communities borne out of the known attributes. Addressing this interplay, we devise a general \emph{block} null model for uncovering unknown community structures, generalized to preserve any form of block structure inherent to the network.

We propose a \emph{block-corrected} modularity and its associated \emph{block} null model for detecting communities in networks with any form of \emph{a priori} known block structure. The probability of an edge from node $i$ to node $j$ under the block null model is
\begin{equation}\label{eq:block-cor}
    P^\mathrm{bc}_{ij} = \frac{k_i^\mathrm{out} k_j^\mathrm{in}}{K_{r}^\mathrm{out} K_{s}^\mathrm{in}} L_{r s},
\end{equation}
where $r$ denotes the \emph{a priori} block of node $i$ and $s$ the block of node $j$, $k_i^\mathrm{in}$ and $k_i^\mathrm{out}$ the in- and out-degree of node $i$, $K_{r}^\mathrm{in}$ and $K_{r}^\mathrm{out}$ the sum of in- and out-degrees of all nodes in block $r$, and $L_{r s}$ the number of out-going edges from block $r$ to $s$. In a citation network, for example, scientific papers are nodes, their publication dates are \emph{a priori} blocks and their citation relationships are embodied through edges.
The block null model preserves, on average, both the in- and out-degree distribution of nodes as well as the edge density between every pair of blocks. The model is generalizable to any existing block structure in networks. Any \emph{a priori} block structure that a network may contain can be captured through $L_{r s}$. Elements of matrix $\mathbf{L}$ are determined from the data, where $L_{r s}$ is the number of edges from block $r$ to $s$.

An effective null model preserves the properties of the observed network and, yet, is divested of any community structure. Retrieving block structures of the known properties (attributes) might not always pass for an interesting question as discussed before. In many applications, the uncovering of communities induced by unknown attributes or properties can make for a compelling find. To this end, the known block structure needs to be discounted by the null model. Such nullification occurs when the \emph{a priori} block structure is captured by the model.

In the absence of such discounting, the associated modularity-based method may be biased by the \emph{a priori} structure, detecting erroneous or suboptimal communities. We illustrate a simple example of this in Sec.~\ref{subsec:intersecting-blocks}.

\section{Illustrative examples}\label{sec:examples}
In what follows, we present two example network models to illustrate (A) how the block-corrected modularity reveals unknown community structures in a simplified network with two node attributes, one known and one unknown, and (B) how it fares under slightly more complex settings where the network is temporal and the connectivity pattern follows specific distributions.

\subsection{Simple network model with hidden communities}\label{subsec:intersecting-blocks}
We devise a simplified example to show the effect of the known and unknown attributes on modularity-based community detection under our block-corrected null model according to Eq.~\eqref{eq:block-cor} and the directed configuration null model~\cite{leicht2008community, molloy1995critical}.

Null model in the directed modularity~\cite{leicht2008community}, is defined as
\begin{equation}\label{eq:dir}
    P^\mathrm{dir}_{ij} = k^\mathrm{out}_i k^\mathrm{in}_j/\sum_i k^\mathrm{out}_i\,,
\end{equation}
where edge probability is proportional to the in-degree and out-degree of nodes, normalized by the total number of edges.
We analytically compute modularity values of specific partitions with respect to both known and unknown structures to assess which partition is preferred by the block-corrected modularity and the directed modularity.

\begin{figure}[!tb]
  \centering
  \includegraphics[width=0.89\linewidth]{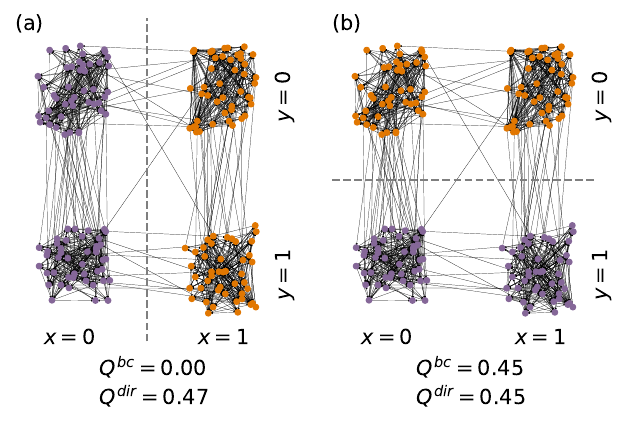}%
  \caption{Network whose nodes are labeled with a known binary attribute $x$ and a unknown binary attribute $y$. Panel (a) shows the partitioning of $x$ and (b) that of $y$, for which modularity values are computed for the block-corrected and the directed models. The network is constructed from 180 nodes equally split between four possible combinations of $x$ and $y$ values, with parameter $p_1^x = 0.6$, $p_1^y = 0.2$, $p_0^x = 0.02$, and $p_0^y = 0.01$.}
  \label{fig:intersecting_schematic}
\end{figure}

Each node has a known binary attribute, $x$, and an unknown binary attribute $y$, taking values zero or one. The probability of connection from node $i$ to $j$ is $p_{ij} = p_{\delta_{x_i x_j}}^{(x)} p_{\delta_{y_i y_j}}^{(y)}$, where $\delta$ denotes the Kronecker delta function. Thus the parameters describing the system are $p_1^{(x)}$, $p_1^{(y)}$, $p_0^{(x)}$, and $p_0^{(y)}$, where the former two affect the connectivity probability of when two nodes have the same values for attributes $x$ and $y$ respectively, and the latter two when values of $x$ and $y$ differ. 
For simplicity, we assume the network is divided into equally sized quadrants of nodes for every possible combination of the $(x, y)$ pair. Figure~\ref{fig:intersecting_schematic} shows a schematic of the network. We refer to this network model as the \emph{intersecting} model in the following discussions.

For such a network, we can define two candidate partitions: one that divides nodes into two classes where $x = 0$ or $x = 1$, whose modularity value is denoted by $Q_x$, and one with two groups of nodes where $y = 0$ and $y = 1$ with its respective modularity value being $Q_y$. Modularity values $Q_y$ and $Q_x$ of a null model are, therefore, quality measures reflecting the superior partition. 
Here, the aim is to find the unknown partition defined by $y$ but masked by the known partition given by $x$. 
To do so, the partitioning across the unknown attribute $y$ needs to be preferred over that across the known attribute $x$, that is, $Q_y$ needs to be larger than $Q_x$.

Deriving the modularity values $Q_x$ and $Q_y$ for our block null model shows that partitioning along the known attribute $x$ results in a negligible modularity gain ($Q^\mathrm{bc}_x \approx 0$ when $|V| \gg 0$, where $Q^\mathrm{bc}_x$ denotes $Q_x$ calculated using the block null model).
For the directed modularity, using the directed null model (Eq.~\eqref{eq:dir}), we likewise compute the modularity value along $x$ to be
\begin{align}\label{eq:Q_x}
    Q^\mathrm{dir}_x &= \frac{1}{2}\frac{p_1^{(x)} - p_0^{(x)}}{p_1^{(x)} + p_0^{(x)}}\,.
\end{align}
Splitting the network with respect to the unknown attribute $y$ yields the same modularity value for both null models:
\begin{equation}\label{eq:Q_y}
   Q^\mathrm{dir}_y = Q^\mathrm{bc}_y = \frac{1}{2}\frac{p_1^{(y)} - p_0^{(y)}}{p_1^{(y)} + p_0^{(y)}}\,.
\end{equation}

The derived modularity expressions for $Q^\mathrm{dir}_y$, $Q^\mathrm{bc}_y$ and $Q^\mathrm{dir}_x$ have a similar form to the E-I index~\cite{krackhardt1988informal} which equates to the number of internal links of group members subtracted from the number of external links divided by their summation. Similar to the E-I index, modularity values $Q_x$ and $Q_y$ are measures of the dominance of in-group over out-group edges. As per Eq.~\eqref{eq:Q_x} and Eq.~\eqref{eq:Q_y} for a given binary attribute the more segregated its two membership groups are, the larger its corresponding modularity value is. The block-corrected modularity evidently rejects the known attribute community structure ($Q^\mathrm{bc}_x = 0$) regardless of how well-separated and embedded its two comprising groups, i.e., the nodes with $x = 0$ and those with $x = 1$, are. In contrast, the directed null model's ability to recover the unknown structure depends on the separatedness of the corresponding attribute's groups. If the separation of the two groups under known attribute $x$ is more pronounced than those of the unknown attribute $y$, the modularity value $Q_x$ is larger than $Q_y$, in which case a modularity optimization partitioning algorithm will dismiss the community structure driven by the unknown attribute $y$ in favor of partitioning the network based on the known attribute $x$. 
The details of the derivations are outlined in Appendix~\ref {appendix:recovering-unknown-block-structure}.

While it may seem plausible to approach this problem by partitioning the network into four communities, i.e., one community for each combination of $x$ and $y$, it is noteworthy that such a partition would only be preferred under certain combinations of parameters $p_0^x$, $p_1^x$, $p_0^y$ and $p_1^y$. In Appendix~\ref{appendix:recovering-unknown-block-structure} we present a combination of parameters where a four-community partition would not be the solution, while the method presented in this paper provides a more general solution for nullifying any pattern of connectivity imposed by known attributes.

\subsection{Communities in temporal networks}\label{sec:temporal-block-structure-ex}
A quintessential example of networks that have directionality with a set of known attributes is citation networks, commonly modeled as directed acyclic graphs (DAGs)~\cite{speidel2015community}. Citation networks are temporal networks in the sense that scientific publications represented by nodes (and citations represented by edges) are timestamped. The timestamps impose structures onto the network that may be classified by community detection methods as genuine communities. In what follows, we use citation networks as our primary class of networks under study.

A null model is characterized by its edge probability $P_{ij}$, which describes a generative network model for networks maximally similar to an observed network with no community structure present. Null models may nonetheless exhibit structures different from those existing in the observed network which can speak to their efficacy as a generative model.
With citation networks in mind, we evaluate the performance of our proposed block null model against two previously proposed null models: the directed null model~\cite{leicht2008community}, which we already introduced in the previous subsection, and the DAG null model~\cite{speidel2015community, karrer2009random}. The directed and the DAG null models are designed to operate on directed networks, with the latter specifically devised with citation networks in mind as they are often represented by directed acyclic graphs\footnote{Real-world citation networks, however, are not always acyclic, requiring pre-processing measures to ensure the acyclic nature holds.}. While the directed null model preserves only in- and out-degrees of nodes, the DAG null model additionally encodes two parameters $\mu_t$ and $\lambda_t$ per time layer $t$ describing the edge connectivity between layers:
\begin{equation}\label{eq:DAG}
P^\mathrm{DAG}_{ij} =
\begin{cases} 
0 & t_i \geq t_j\,, \\
k_i^{\mathrm{in}} k_j^{\mathrm{out}} \frac{\prod_{t = t_i + 1}^{t_j - 1} \lambda_t}{\prod_{t = t_i + 1}^{t_j} \mu_t} & t_i < t_j\,,
\end{cases}
\end{equation}
where 
\begin{equation*}
\mu_t = \sum_{i; t_i < t} k_i^{\mathrm{in}} - \sum_{i; t_i < t} k_i^{\mathrm{out}}
\end{equation*}
and 
\begin{equation*}
\lambda_t = \sum_{i; t_i < t} k_i^{\mathrm{in}} - \sum_{i; t_i \leq t} k_i^{\mathrm{out}}.
\end{equation*}

\subsubsection{Generative network models}\label{sec:generative-models}
We define here a family of generative models for citation networks with specific edge connectivity patterns and timestamped nodes. The realizations of these generative models can be thought of as simplified citation networks, with characteristics of directed acyclic graphs and specified edge length distributions, that is, the time differences of the edge endpoints. The generated networks can thus be used to test the performance of the three definitions of modularity based on different null models.

We assume that each node in the network is assigned two attributes: a discrete timestamp represented by integers $1, \dots, t_\mathrm{max}$, and the membership in a planted community, which is a categorical variable represented by integers $1, \dots, b$. We call the set of nodes that share the same timestamp value a time layer. Each generative model is characterized by three matrices: $\mathbf{N}$ is a $b \times t_\mathrm{max}$ matrix denoting the number of nodes in each planted community in each time layer; $\mathbf{T}$ is a $t_\mathrm{max} \times t_\mathrm{max}$ matrix denoting the effect of time layers on the probability of connectivity; and $\mathbf{B}$ is $b \times b$ matrix specifying the effect of community memberships on the probability of connectivity. Each node $i$ belongs to exactly one planted community $b_i$ and one time layer $t_i$ and the probability of connectivity between nodes $i$ and $j$ is thus simply
\begin{equation}\label{eq:generative_p}
    p_{ij} = B_{b_i b_j} T_{t_i t_j}\,.
\end{equation}

We first define the ``skewed'' variant of our generative model. This variant limits connections to only the temporally adjacent and those in the first and last blocks. 
The probability of two nodes being connected by an edge is greater if both belong to the same planted community, i.e.,
\begin{align}
    N_{b_i t_i} &= N/b, \\
    B_{b_i b_j} &= \begin{cases}
        \frac{\langle k^\mathrm{ing} \rangle}{N/b} &b_i = b_j\,,\\
        \frac{\langle k^\mathrm{outg} \rangle}{N/b} &\text{otherwise},
    \end{cases}\\
    T_{t_i t_j}^\text{skw.} &= \begin{cases}
        1 & t_j = t_i - 1\,,\\
        1 & t_j=1 \text{ and } t_i = t_\mathrm{max}\,,\\
        0 &\text{otherwise}, 
    \end{cases}\label{eq:skewed}
\end{align}
with node rate $N$ determining the number of nodes in each layer\footnote{Node rate $N$ is selected such that it is divisible by the number of planted communities $b$.}, $b$ determining the number of planted communities, and $\langle k^\mathrm{ing} \rangle$ and $\langle k^\mathrm{outg} \rangle$ parametrizing the expected in- and out-group degree of the nodes. 
The distribution of edge length, i.e., the time difference between adjacent nodes, is heavily skewed, with many short edges between adjacent temporal blocks, and a handful of very long ones.

Figure~\ref{fig:grid_block_mats} illustrates a realization of the skewed model. The example synthetic network has $b=2$ planted (horizontal) unknown communities, and \emph{a priori} known (vertical) time layers (or temporal blocks). The connections between the first and the last layer are not visualized.

To contrast this, we construct an ``exponential'' generative model with a probability of connectivity exponentially decaying with edge lengths, i.e.,
\begin{equation}\label{eq:exponential-model}
    T_{t_i t_j}^\text{exp.} = 
    \begin{cases}
    d (1-d)^{t_i - t_j} & \text{for } t_i > t_j\,, \\
    0 & \text{otherwise},
    \end{cases}
\end{equation}
where $d$ is a parameter controlling the rate of decay.

Lastly, we construct a ``power law'' model with a smooth yet heavy-tail probability distribution of edge lengths defined as
\begin{equation}\label{eq:power-law-model}
    T_{t_i t_j}^\text{plw} = 
    \begin{cases}
    \frac{(t_i - t_j)^\gamma}{\zeta(-\gamma)} & \text{for } t_i > t_j\,, \\
    0 & \text{otherwise},
    \end{cases}
\end{equation}
where $\zeta(\cdot)$ is the Riemann zeta function. This provides a more realistic scenario compared to the skewed model, where the temporal edge lengths follow a smoother, heavy-tail probability distribution which, as we will show in Sec.~\ref{sec:openalex}, is corroborated by edge length distributions observed in real-world citation data~\cite{parolo2015attention}.

Note that for all three models, the matrices $\mathbf{N}$ and $\mathbf{B}$ remain the same, and only the connectivity pattern along the known attribute (time) is changed. This allows us to compare the robustness of the three null models with respect to changes in connectivity patterns across time.

\subsubsection{Temporal connectivity pattern in null models}
Each null model, in essence, constructs a function $P_{ij}$ by observing and encoding key values from the given network. These null models can, in turn, function as generative models. Ideally, a null model will reproduce as much of the effects of the known attribute from the original network without reproducing any of the effects of the unknown attribute. This way, the term $\mathbf{A}-\mathbf{P}$ of modularity (Eq.~\eqref{eq:modularity_func}) will reflect the connectivity present (or missing) due to the effect of the unknown attributes.

To assess this, we define a single generalized $(b t_\mathrm{max}) \times (b t_\mathrm{max})$ \emph{connectivity matrix} $\mathbf{M}$ for each temporal network, where $M_{(b t_i + b_i) (b t_j + b_j)}$ is the fraction of existing edges (out of all possible edges) from community $b_i$ at time $t_i$ to community $b_j$ at time $t_j$. We then compare this to the connectivity matrix of the realizations of the same network passed through block-corrected, DAG, and directed null models.

\begin{figure*}[!tb]
  \centering
  \includegraphics[width=\linewidth]{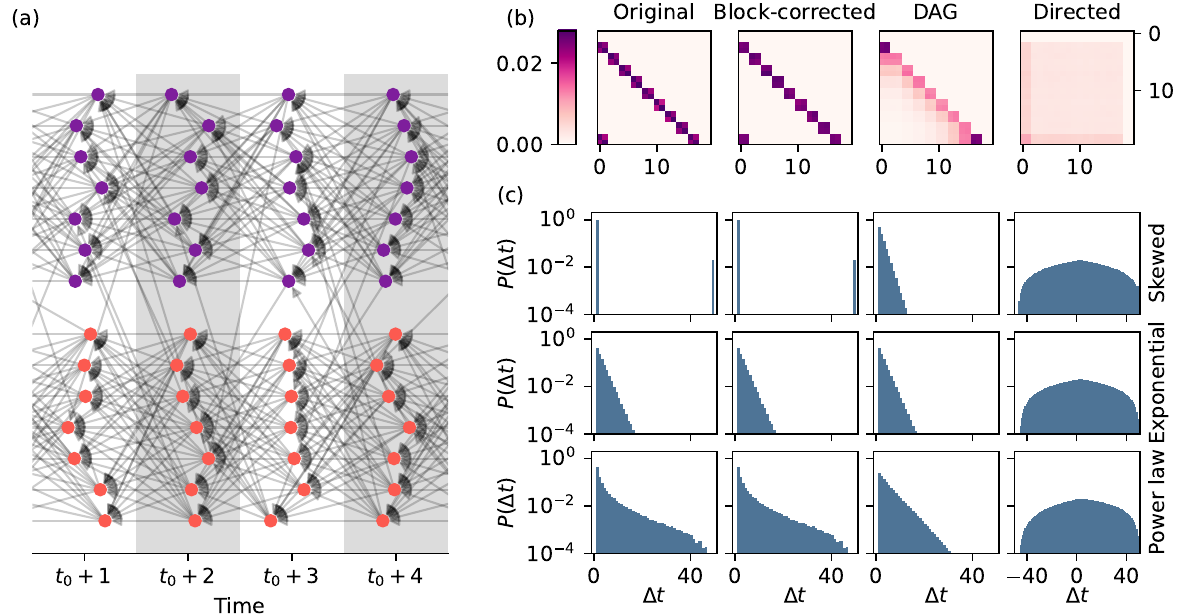}
  \caption{
  (a) Schematic of the skewed generative network model with two planted communities and a temporal structure where only nodes in adjacent blocks are connected. In addition, there are long links from the last block to the first.
  (b) Averaged connectivity matrices $\mathbf{M}$ of an ensemble of 100 network realizations of the block-corrected, DAG, and directed null models. The connectivity matrix of the original network, generated using the skewed model (Eq.~\eqref{eq:skewed}), is labeled ``Original''.
  The skewed model network has $b=2$ planted communities, $t_\mathrm{max} = 10$ temporal layers, $N = 100$ nodes per temporal layer, mean in-group degree $\langle k^\mathrm{ing} \rangle = 10$ and out-group degree $\langle k^\mathrm{outg} \rangle = 8$.
  (c) The distribution of temporal edge lengths for all three network models: one with a skewed distribution of edge lengths (Eq.~\eqref{eq:skewed}), one with an exponential distribution (Eq.~\eqref{eq:exponential-model}), and lastly one with a power law edge length distribution (Eq.~\eqref{eq:power-law-model}). The three observed networks for skewed, exponential, and power law in panel (c) have $t_\mathrm{max} = 50$ temporal blocks, i.e.~time layers, with a fixed number of nodes ($N = 100$) per temporal block. The number of planted communities is $b = 2$ and each node has a mean in-group degree $\langle k^\mathrm{ing} \rangle = 10$ and out-group degree $\langle k^\mathrm{outg} \rangle = 8$. Edge length distributions for the drawn networks from the three null models are shown in their respective columns.}
  \label{fig:grid_block_mats}
\end{figure*}

Figure~\ref{fig:grid_block_mats}(b) depicts the connectivity matrix $\mathbf{M}$ of a network drawn from the skewed network model with two unknown communities, along with connectivity matrices of the three null models. The effect of the known attribute (time) is encoded in the $2\times2$ blocks of nonzero probability of connectivity near the diagonal representing adjacent time layers, as well as the $2\times2$ block at the corner representing connections between the first and the last layer. The structure induced by the unknown attribute is encoded in the difference of colors between diagonal and anti-diagonal elements in each $2\times2$ block.

 Averaged connectivity matrices of the realizations for the block-corrected, DAG, and directed null models are shown in the following three panels. The temporal connectivity patterns produced in our block-corrected model closely resemble those of the original skewed network, whereas the DAG model smooths the temporal connectivity pattern into an exponential decay. The directed model, on the other hand, assumes almost all community-time pairs are equally likely to be connected.

Models' capacity to generate networks close to the observed network is further affirmed by comparing the temporal edge length distribution of the generated ensemble of networks by the null models to that of the observed network. We construct a network with each of the three generative models described in Sec.~\ref{sec:generative-models}, namely, a skewed network (Eq.~\ref{eq:skewed}), a network with exponential edge length distribution (Eq.~\eqref{eq:exponential-model}) with decay $d=0.4$, and a network with power law length distribution (Eq.~\eqref{eq:power-law-model}) with exponent $\gamma = -1.4$. For each network, we draw an ensemble of networks by passing it to each of the three null models. We then compare the edge length distribution of these \emph{nullified} networks to that of the original of the three network models.

Figure~\ref{fig:grid_block_mats}(c) shows that both the block-corrected and the DAG models recover the exponential edge length distribution observed in the original network. This is not the case, however, when the observed edge lengths follow a more heavily skewed distribution. The DAG model falls short of reproducing the distribution, producing an exponential edge length decay instead, while the block-corrected model generates an ensemble whose edge lengths closely follow the observed distribution. Similarly, under a power law distribution for the edge lengths, the DAG model generates networks with exponential edge length distributions instead, whereas the original power law distribution is recreated by the block-corrected model.
In all three generative network models with exponential and skewed distributions, the directed model proves ineffective at capturing the nature of the distributions, producing almost constant temporal edge length distributions only affected by the finite number of temporal blocks $t_\mathrm{max}$.

\subsubsection{Susceptibility to erroneous partitioning}
For an ideal modularity measure, the modularity value of the true partition (the two horizontal color-coded communities of nodes in Fig.~\ref{fig:grid_block_mats}(a)) is larger than that of any other erroneous candidate partition. Otherwise, a modularity maximization algorithm will find the erroneous candidate partition instead of the true partition. To assess whether this holds, for each of the three modularities, we compare the modularity value of the true planted partition $Q_\mathrm{true}$ to that of an erroneous candidate partitioning $Q_\mathrm{cp}$, in which the first and last quarter of all nodes in the temporally ordered network are in one community and the rest in another. This candidate partition places most temporally long edges in one community and the short ones in another. Such partitioning can potentially portray how effective null models are in the face of temporally long edges.

\begin{figure}[!tb]
  \centering
  \includegraphics[width=\linewidth]{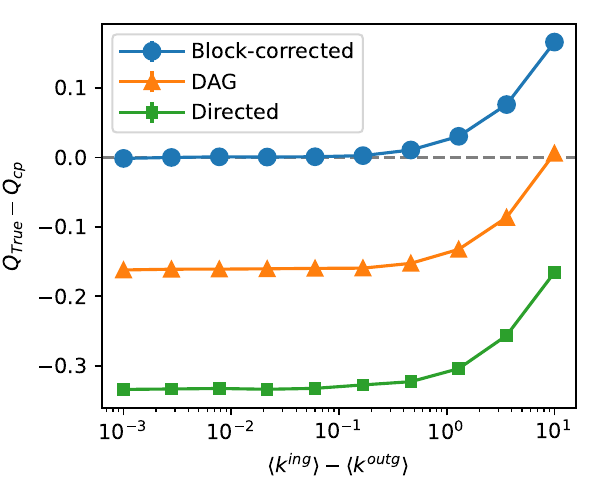}
  \caption{Difference between modularity values for the erroneous candidate partitions, where the first and last $1/4$ of temporal blocks are placed in one community and the rest in another, and that of the true partition over an ensemble of random networks with varying average in-group degrees. All networks are realizations of the skewed model (Eq.~\eqref{eq:skewed}), with $t_\mathrm{max} = 12$ temporal blocks and $N = 100$ nodes per block. $Q_\mathrm{true} > Q_\mathrm{cp}$ indicates that a modularity prefers the true partition to the erroneous candidate. Error bars show 95\% confidence intervals.}
  \label{fig:mod_vals}
\end{figure}

Our results in Fig.~\ref{fig:mod_vals} illustrate that, under the skewed synthetic network model, the block-corrected null model proves to be impervious to the erroneous candidate partitioning ($Q_\mathrm{true} > Q_\mathrm{cp}$). It correctly deems the ground-truth horizontal partition preferable, even in networks whose ground-truth communities are only slightly distinguishable. The other two null models, however, erroneously prefer the incorrect partitioning over the ground-truth communities, even more so in networks whose true communities are not abundantly distinct from each other as quantified by the difference between average in-group and out-group degrees, i.e., as $\langle k^\mathrm{ing} \rangle - \langle k^\mathrm{outg} \rangle$ approaches zero.

As we will describe in Sec.~\ref{sec:openalex}, heavy-tail distributions of edge lengths is observed in certain real-world systems, including in citation networks. As we saw in Fig.~\ref{fig:grid_block_mats}(c), while our block null model can replicate the temporal connectivity pattern induced by a power law edge length distribution, the two other null models do not reproduce the underlying distribution. To understand the implications of this, we directly measure the ability of each modularity definition to recover the planted partitions in the space of all possible bipartitions of the network. To simplify this comparison, we assume the number of planted communities is known in all cases.

The quality of a partition can be quantified by the adjusted Rand score between that and the planted partition. Rand Score in its original form is a similarity measure for two partitionings, defined as
\begin{equation}
    R = \frac{\sum_{i<j}^{|V|} \eta_{ij}}{\binom{|V|}{2}}\,,
\end{equation}
where $\eta_{ij}$ is equal to one if both partitionings agree that $i$ and $j$ are in the same or different communities~\cite{rand1971objective}. Adjusted Rand Score modifies this definition so that a score of zero represents performance no better than random labeling of nodes and one is a complete match between the candidate partitioning and the true one~\cite{hubert1985comparing}:
\begin{equation}
    \frac{R - E[R_\text{random}]}{R_\text{max} - E[R_\text{random}]}\,,
\end{equation}
where $E[R_\text{random}]$ is the expected Rand Score of random labeling and $R_\text{max}$ the maximum possible value for Rand Score corresponding to a perfect match.

\begin{figure*}[!tb]
  \centering
  \includegraphics[width=\linewidth]{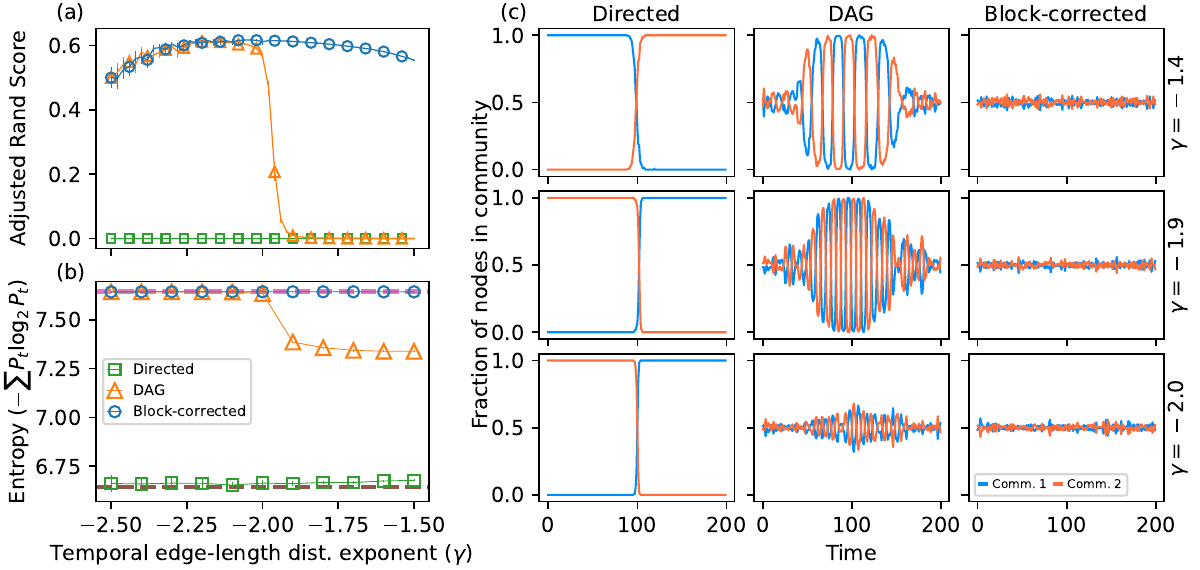}
  \caption{Temporal homogeneity and labeling accuracy for the temporal planted partition model with power law edge length distribution. Adjusted Rand scores between discovered communities and the true partitions (a) and the Shannon entropy of the discovered communities (b) for an ensemble of 50 synthetic networks with two planted communities over a range of exponents of power law edge length distributions (see Eq.~\eqref{eq:power-law-model}).
  Adjusted Rand score of zero indicates the performance is no better than random chance and an adjusted Rand score of 0.6 indicates misplacement of around 11.2\% of $|V|=40\ 000$ nodes. Higher values of entropy indicate more even distribution of communities across time layers.
  The horizontal dashed pink line represents the analytical entropy for a uniform distribution $S=\log_2(t_\mathrm{max})\approx7.64$ while the horizontal dashed brown line corresponds to the analytical entropy for an equal vertical cut $S=\log_2(t_\mathrm{max}/2)\approx6.64$.
  In panel (c) fraction of nodes in each of the two discovered communities is plotted across time layers for three networks with $\gamma = -1.4$, $\gamma = -1.9$, $\gamma = -2.0$.
  The synthetic networks are generated with a fixed number of nodes ($N=200$) per time layer ($t_\mathrm{max} = 200$). Each node has an in-group degree $\langle k^\mathrm{ing} \rangle = 8$ and out-group degree $\langle k^\mathrm{outg} \rangle = 4$. Edges probabilities decrease as $P(\Delta t) \propto \Delta t^{\gamma}$. All three models are coded to find exactly two communities. Error bars in panels (a) and (b) show 95\% confidence interval.}
  \label{fig:grid-power-law-random-dags}
\end{figure*}

Figure~\ref{fig:grid-power-law-random-dags}(a) shows that as the edge length distribution tail becomes heavier, that is, as the absolute value of the exponent $\gamma$ in Eq.~\eqref{eq:power-law-model} becomes smaller, the block-corrected modularity retains its ability to distinguish the ground-truth partition, whereas a sharp drop is observed in the case of the DAG modularity. The directed modularity fails to capture the planted temporal community structure in the networks altogether and its adjusted Rand score remains at zero across all exponents. The sharp drop in performance of the DAG modularity, ultimately, stems from the fact that its implicit connectivity assumption, i.e., exponential edge length distribution, does not match the connectivity patterns present in the realized networks.

In the \emph{intersecting} network model we studied in Sec.~\ref{subsec:intersecting-blocks}, where nodes come with exactly one \emph{a priori} and one unknown attribute, we observed how the directed modularity is likely to divide the network with respect to the known attribute. Under the \emph{power law} random network model where node timestamps operate as the \emph{a priori} or known attribute, it stands to reason to speculate that the low adjusted Rand scores for directed modularity, resembling those of close-to-random labeling, are rooted in the network being divided into two temporally separated clusters.

In such a case, community entropies quantifying how uniformly each community is distributed across time layers should be relatively small. For the same ensemble of networks in Fig.~\ref{fig:grid-power-law-random-dags}(a), we plot the entropy of one of the two discovered communities over a range of power law exponents. For a community, the probability $P_t$ that a randomly selected node belongs to time layer $t$ is computed and the entropy associated with the community is thus $-\sum_t P_t\log_2 P_t$. As evidenced by Fig.~\ref{fig:grid-power-law-random-dags}(b) the entropies for the directed modularity are notably smaller than that of the block-corrected and the DAG modularities, remaining close to the entropy for equal vertical cut $S=\log_2(t_\mathrm{max}/2)\approx6.64$ for $t_\mathrm{max} = 200$. Entropy of the block-corrected modularity keeps at the analytical entropy for a uniform distribution $S=\log_2(t_\mathrm{max})\approx7.64$ for all values of exponents.
It is noteworthy that the block-corrected and the DAG modularities begin to diverge in entropy values starting from the same threshold of $\gamma \geq -2$ as in Fig.~\ref{fig:grid-power-law-random-dags}(a) where the adjusted Rand score plummets for the DAG model. That is to say, as the edge length distribution becomes more heavy-tailed, the entropy level falls for the DAG modularity, while the block-corrected modularity remains unaffected.

This hypothesized vertical cut preferred by the directed modularity, dividing the network into two communities of older and younger nodes, is further substantiated by Fig.~\ref{fig:grid-power-law-random-dags}(c) which illustrates the fraction of nodes in the two discovered communities over the time layers. Three networks are realized from the aforementioned power law generative network model (see Eq.~\eqref{eq:power-law-model}) each parametrized by exponent $\gamma$.
With the block-corrected modularity, nodes in both communities are evenly distributed across time layers for all values of $\gamma$, as corroborated by the entropy value continuously remaining close to the analytical value for a homogeneous distribution of the community nodes across time as seen in Fig.~\ref{fig:grid-power-law-random-dags}(b). This more closely resembles the distribution of the nodes in the planted communities, where there are $N/2$ nodes of each community in each layer.

The DAG modularity, however, alternatively assigns the majority of the nodes of groups of time layers to one of the two communities for smaller values of $\gamma$. This behavior, which is not reminiscent of the planted (original) communities, can already be seen at $\gamma = -2.0$ and seems to aggravate for more heavy-tailed distributions. This, again, corroborates the observations of a drop in entropy in Fig.~\ref{fig:grid-power-law-random-dags}(b).
As hypothesized before, the directed modularity makes a vertical cut for all three values of $\gamma$ visualized in Fig.~\ref{fig:grid-power-law-random-dags}(c). The cut seems to become more gradual as the magnitude gets smaller, i.e., the heavy tail characteristic of the edge length distribution augments. 

\section{Heuristics for modularity maximization}\label{sec:heuristics}
Maximizing the modularity function over all possible partitionings of the network is computationally hard, an NP-hard problem~\cite{brandes2007modularity}, hence the development of an array of heuristic algorithms~\cite{clauset2004finding, blondel2008fast}. Here, we opt for the spectral method~\cite{newman2006finding}, the so-called leading eigenvector method. Initially designed for finding exactly two groups of nodes in the network in the simplest setting, spectral bipartitioning can easily be extended to an arbitrary number of communities by repeatedly dividing the communities into two until no modularity gain is to be achieved by so doing.

While in many scenarios the leading eigenvector produces satisfactory outcomes~\cite{leicht2008community}, the resulting partition can nonetheless be improved by reassigning nodes to alternate communities. This has prompted the application of fine-tuning steps in many implementations of the spectral method~\cite{leicht2008community, speidel2015community} where nodes are moved among communities for possible gains in the modularity value. In Sec.~\ref{subsec:louvain-fine-tuning} and Sec.~\ref{subsec:split-fine-tuning} we present two slightly distinct, commonly used greedy fine-tuning algorithms, along with their vectorized implementation optimized for sparse representations.

\subsection{Spectral partitioning}\label{subsec:vectorised-px}
Given a network, finding two communities is equivalent to discovering a membership vector, $s$, whose elements are community labels of $+1$ and $-1$ of all nodes, $\mathbf{s} = (s_1, \ldots, s_{|V|})^\intercal$, $s_i \in \{+1, -1\}$. Modularity value in Eq.~\eqref{eq:modularity_func} can be thus re-written as a function of membership vector $\mathbf{s}$ and modularity matrix $\mathbf{B}$, whose element $B_{ij}$ is equal to $A_{ij} - P_{ij}$, and 
\begin{equation} \label{eq:mod_value_matrix_form}
    Q = \frac{1}{2m} \mathbf{s}^\intercal \mathbf{B} \mathbf{s}\,.
\end{equation}

Note that if rows and columns of $\mathbf{B}$ do not sum to zero, modularity value $Q$ for an undivided network would not be zero. In such a case sum of rows must be subtracted from the diagonal of $\mathbf{B}$~\cite{newman2006modularity}.

An approximate solution to this bipartitioning problem can be found by relaxing the integral constraint $s_i \in \{-1, +1\}$ to the real domain \cite{newman2013spectral}. An effective heuristic for the problem is setting the values of $s_i$ based on the leading eigenvector of $\mathbf{B}$ \cite{leicht2008community}, that is, $s_i = \mathrm{sgn}(u_i) \text{, } \forall i \in [n]$, where $\mathbf{u}$ is the leading eigenvector of $\mathbf{B}$.

The network is divided into two communities $\{C_1, C_2\}$ as above and subsequently, in an iterative fashion, the subgraph induced by community $C \in \{C_1, C_2\}$ is then bipartitioned into two smaller communities. 
The contribution to the modularity value of breaking up $C$ into two partitions can be calculated as
\begin{equation}\label{eq:dq}
    \Delta Q = \frac{1}{2m} \Tilde{\mathbf{s}}^\intercal \Tilde{\mathbf{B}}^{(C)} \Tilde{\mathbf{s}}\,,
\end{equation}
where $\tilde{\mathbf{s}}$ is the membership vector of community $C$ to be solved, and $\tilde{\mathbf{B}}^{(C)}$ a generalized modularity matrix of size $|C| \times |C|$ where $\tilde{B}_{ij}^{(C)} = B_{ij} - \delta_{ij}\sum_{k \in C} B_{ik}$ for $i, j \in C$ \cite{speidel2015community, newman2006finding}. From this point forward the superscript $(C)$ for a matrix denotes a submatrix containing rows and columns in $C$. The bipartition of community $C$ is executed only when the corresponding $\Delta Q$ is positive, indicating that performing the bipartitioning results in a gain in modularity value. The iterative bipartitioning halts when modularity value $Q$ stops increasing, meaning $\Delta Q$ is non-positive. In the very first step of the algorithm, community $C$ corresponds to the entire network ($C = V$) in which case, assuming no self-loops are present, $\tilde{\mathbf{B}}^{(V)} = \mathbf{B}$ and $\tilde{\mathbf{s}} = \mathbf{s}$, resulting in $\Delta Q$ in Eq.~\eqref{eq:dq} being equal to $Q$ in Eq.~\eqref{eq:mod_value_matrix_form}. 

In the more general case of directed networks, $\tilde{\mathbf{B}}^{(C)}$ is not necessarily symmetric. This asymmetry might cause issues when finding the leading eigenvector~\cite{leicht2008community}. This can be resolved by restoring symmetry to the $\tilde{\mathbf{B}}^{(C)}$ by adding to it its transpose~\cite{leicht2008community, speidel2015community}. Therefore $\Delta Q$ becomes
\begin{equation*}
        \Delta Q = \frac{1}{4m} \tilde{\mathbf{s}}^\intercal \left(
        \tilde{\mathbf{B}}^{(C)} + ({\tilde{\mathbf{B}}^{(C)}})^\intercal
        \right) \tilde{\mathbf{s}}\,.
\end{equation*}

Computing the leading eigenvector of matrix $\tilde{\mathbf{B}}^{(C)} +({\tilde{\mathbf{B}}^{(C)}})^\intercal$ can be done efficiently via the power method which amounts to iteratively calculating
\begin{equation}\label{eq:iterative-Bx}
    \mathbf{x}^{(i+1)} = \left(\tilde{\mathbf{B}}^{(C)} +({\tilde{\mathbf{B}}^{(C)}})^\intercal\right)(\mathbf{x}^{(i)}/\lvert \mathbf{x}^{(i)}\rvert)
\end{equation}
in every step $i$ starting with an arbitrary vector $\mathbf{x}^{(0)}$ in the first step $i = 0$, until $\mathbf{x}^{(i)}$ converges to the leading eigenvector. 

Modularity matrix is, however, dense and, for example, the explicit matrix-vector multiplication $\mathbf{B}\mathbf{x}$ requires $O(|V|^2)$ operations. Ref~\cite{newman2006finding} shows how the matrix-vector multiplication $\mathbf{B}\mathbf{x}$ with a simpler definition of modularity can be rewritten in vector notation to allow for fast computation. This derivation is not directly applicable to our block-corrected modularity, and we next derive a similar procedure for it. We can take advantage of the definition of $B$ and the fact that typically adjacency matrix $\mathbf{A}$ is sparse to write it as $\mathbf{B}\mathbf{x} = \mathbf{A}\mathbf{x} - \mathbf{P}\mathbf{x}$.
Recalling edge probability in Eq.~\eqref{eq:block-cor}, we define 
\begin{equation*}
    \kappa_i^\mathrm{out/in} = \frac{k_i^\mathrm{out/in}}{K_{t_i}^\mathrm{out/in}}
\end{equation*}
and write out element $i$ of $\mathbf{P}\mathbf{x}$ as 
\begin{align*}
(&\mathbf{P} \mathbf{x})_i = \sum_{j = 1}^n P_{ij}x_j = \sum_{j = 1}^n \kappa_j^\mathrm{in} \kappa_i^\mathrm{out} L_{t_i t_j} x_j \\
&= \kappa_i^\mathrm{out} \sum_{t = 1}^{t_\mathrm{max}} L_{t_i t} \sum_{j \in t} \kappa_j^\mathrm{in} x_j \\
&= \kappa_i^\mathrm{out} \sum_{t = 1}^{t_\mathrm{max}} L_{t_i t} \langle \kappa^{\mathrm{in}(t)}, x^{(t)}\rangle\,.
\end{align*}
Writing this in terms of the element-wise multiplication operator $\odot$ gives
\begin{equation}\label{eq:px-vectorised}
\mathbf{P} \mathbf{x} = \mathbf{\kappa}^{out} \odot (\mathbf{L} \mathbf{v}^\mathrm{in})_{t_\mathrm{max}\times 1 \mapsto \lvert V \rvert \times 1}\,,
\end{equation}
where $v^\mathrm{in}_t = \langle \mathbf{\kappa}^{\mathrm{in}(t)}, \mathbf{x}^{(t)} \rangle $. Here, $(\mathbf{L} \mathbf{v}^\mathrm{in})_{t_\mathrm{max} \times 1 \mapsto \lvert V \rvert \times 1}$ denotes vector $\mathbf{L} \mathbf{v}^\mathrm{in}$ extended to $\lvert V \rvert$ elements by repeating each value in proportion to the number of nodes in the corresponding block. This can only be done if the nodes are sorted by blocks. If not, the nodes are relabeled to satisfy the condition.

Matrix $\mathbf{L}$ is of dimension $t_\mathrm{max} \times t_\mathrm{max}$, with $t_\mathrm{max}$ denoting the number of blocks in the network, and $\mathbf{v}^\mathrm{in}$ is a $t_\mathrm{max} \times 1$ vector. Multiplication $\mathbf{L} \mathbf{v}^\mathrm{in}$ is easily mapped from a $t_\mathrm{max}$-dimensional space to an $\lvert V \rvert$-dimensional one since all nodes in block $t$ would have the same values for both $\mathbf{L}$ and $\mathbf{v}^\mathrm{in}$.
Calculating $\mathbf{B}\mathbf{x}$ using sparse matrix operations therefore has an overall time complexity of $O(m + \lvert V \rvert + t_\mathrm{max}^2)$.

This formulation, however, is only useful as it is for the first bipartitioning of the network. In the general case expressed in Eq.~\eqref{eq:iterative-Bx}, we need to calculate $\tilde{\mathbf{B}}^{(C)} \mathbf{x}$ and $(\tilde{\mathbf{B}}^{(C)})^\intercal \mathbf{x}$ for any $C \subseteq V$ in a vectorized manner. As a first step, we can calculate $\mathbf{B}^{(C)}\mathbf{x} = \mathbf{A}^{(C)}\mathbf{x} - \mathbf{P}^{(C)}\mathbf{x}$. We can carefully modify Eq.~\eqref{eq:px-vectorised} to derive the second term $\mathbf{P}^{(C)}\mathbf{x}$ by (i) subsetting occurrences of $\kappa^\mathrm{in/out}$ to only include nodes in $C$, and (ii) extending term $(\mathbf{L} \mathbf{v}^\mathrm{in})$ from $t_max \times 1$ to $\lvert C \rvert \times 1$ instead of $\lvert C \rvert \times 1$.

Based on these building blocks, we calculate a vectorised notation for $\tilde{\mathbf{B}}^{(C)}\mathbf{x}$. The value for $\tilde{\mathbf{B}}^{(C)}\mathbf{x}$ can be calculated by decomposing it as
\begin{equation}
    \tilde{\mathbf{B}}^{(C)}\mathbf{x} = \mathbf{B}^{(C)}\mathbf{x} - (\mathbf{B}^{(C)}\vec{\mathbf{1}}) \odot \mathbf{x}\,,
\end{equation}
where $\vec{1}$ is a column vector of ones. Vector $\mathbf{B}^{(C)}\vec{1}$, whose $i$-th element is simply the sum of $i$-th row of $\mathbf{B}^{(C)}$, can be efficiently calculated using the same method described in Eq.~\eqref{eq:px-vectorised}. This results in
\begin{equation}\label{eq:bx-opened}
    \tilde{\mathbf{B}}^{(C)} \mathbf{x} = \mathbf{A}^{(C)}\mathbf{x} - \mathbf{A}^{(C)}\vec{\mathbf{1}} \odot \mathbf{x} - \mathbf{P}^{(C)}\mathbf{x} + \mathbf{P}^{(C)}\vec{\mathbf{1}} \odot \mathbf{x}\,.
\end{equation}

$\tilde{\mathbf{B}}^\intercal\mathbf{x}$ can be similarly calculated through a substitution of $\tilde{\mathbf{B}}$ with $\tilde{\mathbf{B}}^\intercal$ in Eq.~\eqref{eq:bx-opened}. Notably, following the same pattern as derivation of Eq.~\eqref{eq:px-vectorised}, for $\mathbf{P}^\intercal \mathbf{x}$ we get
\begin{equation}
    \mathbf{P}^\intercal \mathbf{x} = \mathbf{\kappa}^{in} \odot (\mathbf{L}^\intercal \mathbf{v}^{out})_{t_\mathrm{max} \times 1 \mapsto n \times 1}\,.
\end{equation}

Note that the power iteration method as described converges to the eigenvector corresponding to the largest eigenvalue by absolute value, while the bipartitioning heuristic requires finding the sign of the eigenvector corresponding to the largest eigenvalue. If the largest eigenvalue by absolute value ($\lambda$) returned by the power iteration method is negative, the power iteration process can be repeated for the matrix $\tilde{\mathbf{B}}^{(C)} + (\tilde{\mathbf{B}}^{(C)})^\intercal - \lambda \mathbf{I}$ to find the latter.

\subsection{Split-Louvain fine-tuning}\label{subsec:split-fine-tuning}
The algorithm, based on the implementation in Ref.~\cite{speidel2015community}, can optimize the outcome of a single bipartitioning step from an initial heuristic by calculating the effect of moving each node to the opposite community on modularity and performing the most beneficial move. Our implementation enhances the overall performance by minimizing the amount of redundant calculations and maximizing the use of vectorised operations.

Every bipartition is characterized by membership vector $\mathbf{s}$ where $\mathbf{s}_i \in \{+1, -1\}$. The change in $Q$ for a hypothetical change in community membership of node $k$ we calculate to be $(-4s_k(\mathbf{\Tilde{B}} + \mathbf{\Tilde{B}}^\intercal)_k \mathbf{s} + 4(\mathbf{\Tilde{B}} + \mathbf{\Tilde{B}}^\intercal)_{kk})/4m$. This can be derived by calculating the modularity for the new membership vector $\mathbf{s}' = \mathbf{s} - 2 s_k \mathbf{e}_k$ where $\mathbf{e}_k$ is the $k$-th standard basis vector.

Note that, for readability we will describe the derivations in this section for the case of bipartitioning the entire network ($C$ = $V$). For the more general case of fine-tuning a bipartition of $C \subseteq V$, we can simply replace $\mathbf{s}$ with $\tilde{\mathbf{s}}$ and $\tilde{\mathbf{B}}$ with $\tilde{\mathbf{B}}^{(C)}$.

In a similar fashion, we can compute a vector of all possible improvements $\mathbf{d}$ where $d_k$ is the change in $\Delta Q$ for the hypothetical movement of node $k$ to the other community:
\begin{equation}\label{eq:improvements}
    \mathbf{d} = \frac{1}{m} \left[
        \mathrm{diag}(\tilde{\mathbf{B}} + \tilde{\mathbf{B}}^\intercal)
        -\mathbf{s}\odot(\tilde{\mathbf{B}} + \tilde{\mathbf{B}}^\intercal)\mathbf{s}
    \right]\,,
\end{equation}
where $\odot$ indicates the Hadamard product. $(\tilde{\mathbf{B}} + \tilde{\mathbf{B}})\mathbf{s}$ can be efficiently computed using the sparse matrix operations discussed in Sec.~\ref{subsec:vectorised-px}.

Upon selecting the node with the largest value in vector $\mathbf{d}$ ($k = \argmax{\mathbf{d}}$), a running variable for the accumulation of all $\Delta Q$ values is incremented by $d_k$, after which the improvements vector $\mathbf{d}$ is recomputed and the selecting of a node is repeated for the updated improvements vector $\mathbf{d}$. This is repeated until no modularity-increasing moves remain on unmoved nodes.
For compatibility reasons with the implementation in Ref~\cite{speidel2015community}, we limit the maximum number of movements for every node to one. 

We find it more efficient to apply the effect of changing the membership of node $k$ directly to $\mathbf{d}$ instead of re-computing the entire vector $\mathbf{d}$ from scratch in every iteration:
\begin{equation}
    \mathbf{d'} = \mathbf{d} + \frac{2}{m} \mathbf{s}_k \mathbf{s} \odot (\tilde{\mathbf{B}} + \tilde{\mathbf{B}}^\intercal)_k\,,
\end{equation}
with the exception element $k$ where $\mathbf{d'}_k = -\mathbf{d}_k$.
This computation avoids the repetition of the sparse matrix-vector product in Eq.~\ref{eq:improvements}.

\subsection{Final-Louvain fine-tuning}\label{subsec:louvain-fine-tuning}
In addition to the split-Louvain fine-tuning, we implement another method of fine-tuning akin to phase one of the Louvain algorithm \cite{blondel2008fast}, that is applicable to a partition with an arbitrary number of communities. Unlike the split-Louvain fine-tuning nodes can be moved between communities multiple times, so long as the corresponding $\Delta Q$ values are positive. While in split-Louvain fine-tuning, the best possible move among all nodes is selected in each iteration, in the multi-community final-Louvain fine-tuning, we perform the most beneficial move for every node in an arbitrary order of nodes in each round and repeat it until no modularity-gaining move remains for any of the nodes.
Unlike the split-Louvain fine-tuning which is applied after every bipartition, we only perform the final-Louvain fine-tuning on the final output partition of the community detection algorithm. 

In order to leverage the sparse matrix operations outlined in Sec.~\ref{subsec:vectorised-px}, we define matrix $\mathbf{D}$ of dimension $n \times b$, $n$ being the number of nodes and $b$ the number of communities in the partition, where element $D_{kc}$ is defined to be the modularity contribution of node $k$ to community $c$. This is by definition equal to $(\tilde{\mathbf{B}} + \tilde{\mathbf{B}}^\intercal)\mathbf{G}$ where $\mathbf{G} \in \{0, 1\}^{n \times b}$ denotes the community-membership matrix whose element $G_{kc}$ is one if node $k$ belongs to community $c$ and zero otherwise. Column $j$ in $\mathbf{D}$ is denoted by $\mathbf{D}_{*j}$ that equates $(\tilde{\mathbf{B}} + \tilde{\mathbf{B}}^\intercal)\mathbf{G}_{*j}$ where likewise $\mathbf{G}_{*j}$ is column $j$ of matrix $\mathbf{G}$, and this matrix-vector multiplication can be computed efficiently as described in Sec.~\ref{subsec:vectorised-px}. Thus, this allows for an efficient populating of matrix $\mathbf{D}$.

To decide whether node $k$ be moved to a different community, a $\Delta Q_k$ is computed, that is the gain in modularity when detaching node $k$ from its original community $b_k$, subtracted from that of moving it to a candidate community $c$ where the gain is maximum across all possible communities. These two quantities are immediately accessed via matrix $\mathbf{D}$:
\begin{equation}
    \Delta Q_k = D_{k c} - D_{k b_k}.
\end{equation}

Node $k$ is moved to community $c$ if this value is larger than zero. In the case of a move, the $k$-th element of the respective membership vectors in $\mathbf{G}$ are updated accordingly, and so is matrix $\mathbf{D}$:
\begin{align*}
    D_{k b_i} = D_{k b_i} - (\tilde{\mathbf{B}} + \tilde{\mathbf{B}}^\intercal)_{*k} \\
    D_{k c} = D_{k c} + (\tilde{\mathbf{B}} + \tilde{\mathbf{B}}^\intercal)_{*k}. \\
\end{align*}

This update rule allows for an efficient re-computation of the changed columns of $\mathbf{D}$ without having to re-calculate the entire matrix.

\section{Real-world Citation Data: OpenAlex}\label{sec:openalex}
To assess the performance and scalability of our modularity method, we apply it to a set of large real-world citation networks varying in size and academic field.
For our purposes, we rely on the extensive OpenAlex citation data set~\cite{priem2022openalex}. Publications in the OpenAlex database are accompanied by their scientific fields, with different levels of granularity, ranging from topic and subfield to scientific domain, as well as information about publication venues, author affiliations, and funding sources. 

\subsection{Field networks}
\begin{table*}
\caption{Community detection results of the block-corrected modularity-based algorithm on a list of OpenAlex networks, along with their respective sizes and academic fields. ``Duration'' specifies running time, ``Comms'' specifies the total number of communities and the smallest number of communities that contain at least 90\% of the nodes in parentheses. The split-Louvain and final-Louvain are abbreviated as sLouv and fLouv respectively.
Results are averaged on an ensemble of 100 independent runs.}\label{tab:OA-field-networks}
    \centering\footnotesize
\begin{tabular}{l l c c c c c c c c}
 &  & Veterinary & Dentistry & Energy & Economics & Psychology & Neuroscience & Phys. \& Astro. & CS\\
\hline\hline
Nodes &  & 50.6K & 126.8K & 311.6K & 353.1K & 781.3K & 890.1K & 1.1M & 1.2M\\
Edges &  & 470.5K & 1.7M & 7.9M & 4.1M & 12.6M & 21.0M & 23.1M & 14.0M\\
\hline
\multirow{3}{*}{Comms.} & No finetune & 7.2 & 5.5 & 8.6 & 37 & 12 & 13 & 20 & 14\\
 & sLouv & 53 & 44 & 25 & 46 & 38 & 31 & 40 & 58\\
 & fLouv & 6.9 & 5.5 & 8.5 & 36 & 13 & 14 & 20 & 16\\
 & Both & 51 & 42 & 24 & 46 & 36 & 30 & 41 & 59\\
\hline
\multirow{3}{*}{\makecell{Large\\Comms.}} & No finetune & 3.1 & 1.0 & 1.0 & 15 & 2.5 & 5.8 & 6.1 & 5.1\\
 & sLouv & 24 & 15 & 8.8 & 15 & 16 & 12 & 18 & 21\\
 & fLouv & 3.3 & 3.6 & 4.8 & 13 & 5.3 & 7.5 & 9.4 & 7.3\\
 & Both & 23 & 16 & 8.4 & 13 & 15 & 13 & 18 & 21\\
\hline
\multirow{3}{*}{$Q$} & No finetune & 0.487 & 0.369 & 0.346 & 0.228 & 0.287 & 0.327 & 0.394 & 0.228\\
 & sLouv & 0.824 & 0.729 & 0.682 & 0.657 & 0.659 & 0.64 & 0.791 & 0.744\\
 & fLouv & 0.659 & 0.661 & 0.632 & 0.657 & 0.62 & 0.638 & 0.73 & 0.641\\
 & Both & 0.834 & 0.751 & 0.699 & 0.706 & 0.692 & 0.671 & 0.807 & 0.769\\
\hline
\multirow{3}{*}{Duration} & No finetune & 93s & 66s & 8.6m & 30m & 30m & 39m & 2.0h & 57m\\
 & sLouv & 14m & 19m & 37m & 58m & 2.7h & 3.0h & 6.5h & 7.6h\\
 & fLouv & 1.9m & 3.3m & 19m & 70m & 3.7h & 3.5h & 6.6h & 6.5h\\
 & Both & 13m & 20m & 38m & 62m & 3.5h & 4.2h & 7.4h & 9.8h\\
 \end{tabular}
\end{table*}

We select eight networks ranging in size and scientific discipline as summarized in Table~\ref{tab:OA-field-networks}. Nodes in the networks are articles whose publication dates span from the beginning of the year 2000 to 2025, removing all whose in-degrees are less than 10. We bin nodes in time by year, resulting in 24 time layers in total. Since the networks need not be directed acyclic, we allow for edges within time layers or against the arrow of time. 

For each network, we run our community detection method and report the mean and the standard error of the mean of the number discovered of communities (comms.), the number of discovered communities comprising in total more than 90\% of the network (Large comms.), modularity value ($Q$), and runtime (Duration) across 50 runs. We execute additional fine-tuning steps in two slightly different forms, namely split-Louvain and final-Louvain. Split-Louvain fine-tuning is performed following every bipartition step and is akin to that performed in~\cite{newman2006modularity, leicht2008community}. Final-Louvain fine-tuning is the multi-community counterpart to split-Louvain fine-tuning applied to the discovered partition by the community detection method, resembling the first phase of the Louvain algorithm ~\cite{blondel2008fast}. To achieve improved runtime we leverage the vectorised matrix operations described in Sec.~\ref{subsec:vectorised-px} in computing the changes in modularity values for both fine-tuning processes, the details of which we presented in Sec.~\ref{subsec:split-fine-tuning} and Sec.~\ref{subsec:louvain-fine-tuning}.

Our implementation of spectral partitioning, detailed in~\ref{subsec:vectorised-px}, involves computing leading eigenvectors by way of the power iteration method, parameterized by the tolerance within which the leading eigenvalue is converged. Tolerance affects the runtime and dictates the precision of the converged eigenvector. Producing Table~\ref{tab:OA-field-networks} we observe that a tolerance of $10^{-10}$ is a reasonable enough choice, yielding eigenvalues containing six to eight significant decimal digits on average across all runs.
There are limitations imposed by the tolerance parameter that account for the non-zero errors of the number of communities in Table~\ref{tab:OA-field-networks}. What we observe first-hand on both empirical and synthetic networks is the existence of nodes whose values in the leading eigenvector are zero or notably small. The signs of these values can flip across runs, resulting in different group assignments for nodes, and subsequently different partitionings of the network altogether. As the spectral partitioning process is iterative, these alternative group assignments can compound over subsequent iterations, resulting in larger deviations.

As we illustrate in Appendix~\ref{appendix:tol}, the tolerance value required for achieving a specific level of accuracy in a single bipartitioning step is intrinsically linked with the network structure. We show empirically, using synthetic network models, that the required tolerance scales with network properties such as the number of temporal layers as well as the average temporal edge lengths. Such dependence on the network structure, especially in empirical settings where errors can accumulate over the many iteration steps of the spectral partitioning, can incur errors on the final output and therefore lead to fluctuations across runs.

For all networks in Table~\ref{tab:OA-field-networks}, the split-Louvain fine-tuning results in larger modularity values than the final-Louvain. One way to explain this is by the same phenomenon of error accumulation that, we posit, accounts for the fluctuations across community detection outcomes of a network. Split-Louvain fine-tuning is executed after every bipartition and thus diminishes the error that might compound over all subsequent iterative steps. In contrast, final-Louvain fine-tuning is performed only on the final outcome partition. Although applying both types of fine-tuning only marginally improves the results of split-Louvain, it typically adds a negligible amount of extra computation time compared to using split-Louvain alone.

Also notable in the result is that the application of the split-Louvain method (or the simultaneous application of both fine-tuning methods) often results in a larger number of detected communities, as well as a notable increase in the number of large detected communities, whereas the spectral method often results in only one or very few large communities. This, combined with the noteworthy gain in modularity, suggests that an effective fine-tuning step is often necessary to alleviate the shortcomings of the spectral method.

\subsection{Subfield networks and metadata alignment evaluation}
\begin{table*}[htb!]
\caption{Community detection results of the block-corrected modularity-based algorithm on a list of OpenAlex networks consisting of two subfields. Results are averaged on an ensemble of 100 independent runs.}\label{tab:OA-subfields-no-init}
    \centering\footnotesize
\begin{tabular*}{\textwidth}{@{\extracolsep{\fill}} l l c c c c} 
  &  & \makecell{Biochem. --\\Clinical biochem.} & \makecell{Environmental chem. --\\Analytical chem.} & \makecell{Environmental chem. --\\Ocean eng.} & \makecell{Environmental eng. --\\Pollution}\\
\hline\hline
Nodes &  & 73.8K & 179.3K & 209.2K & 333.2K\\
Edges &  & 951.6K & 2.6M & 2.6M & 5.4M\\
Metadata Q &  & 0.428 & 0.451 & 0.432 & 0.420\\
\hline
\multirow{3}{*}{$Q$} & No finetune & 0.186 & 0.140 & 0.138 & 0.143\\
 & sLouv & 0.429 & 0.434 & 0.438 & 0.462\\
 & fLouv & 0.450 & 0.437 & 0.450 & 0.453\\
\hline
\multirow{3}{*}{$Q_\text{randomized}$} & No finetune & 0.048 & 0.032 & 0.032 & 0.026\\
 & sLouv & 0.118 & 0.113 & 0.120 & 0.106\\
 & fLouv & 0.120 & 0.115 & 0.121 & 0.107\\
\hline
\multirow{3}{*}{$Duration$} & No finetune & 3.9s & 14s & 14s & 32s\\
 & sLouv & 29s & 5.4m & 5.3m & 21m\\
 & fLouv & 51s & 4.5m & 9m & 19m\\
    \end{tabular*}
\vspace*{1em}

\begin{tabular*}{\textwidth}{@{\extracolsep{\fill}} l l c c c c}
 &  & \makecell{Pharmacology --\\Social psych.} & \makecell{Polymers \& plastics --\\Biomater.} & \makecell{Aerosp. eng. --\\Mech. of material} & \makecell{Physiology --\\Cardio. med.}\\
\hline\hline
Nodes &  & 389.2K & 407.5K & 491.2K & 772.0K\\
Edges &  & 5.1M & 7.9M & 5.4M & 13.4M\\
Metadata Q &  & 0.461 & 0.370 & 0.427 & 0.417\\
\hline
\multirow{3}{*}{$Q$} & No finetune & 0.165 & 0.160 & 0.0591 & 0.287\\
 & sLouv & 0.440 & 0.445 & 0.457 & 0.449\\
 & fLouv & 0.468 & 0.417 & 0.449 & 0.454\\
\hline
\multirow{3}{*}{$Q_\text{randomized}$} & No finetune & 0.027 & 0.033 & 0.025 & 0.022\\
 & sLouv & 0.117 & 0.097 & 0.128 & 0.102\\
 & fLouv & 0.118 & 0.099 & 0.129 & 0.103\\
\hline
\multirow{3}{*}{$Duration$} & No finetune & 32s & 43s & 35s & 1.8m\\
 & sLouv & 27m & 28m & 42m & 90m\\
 & fLouv & 25m & 25m & 37m & 56m\\
 \end{tabular*}
\end{table*}

In a similar fashion to field networks, subfield networks can be constructed, containing scientific works sharing one or more closely related subfields. Subfields are the second-most granular level of the ``domain, field, subfield, topic'' system that tags publications. Limiting the number of subfields to two, we build networks where only a single bipartition can not only evade the complexities of the recursive spectral bipartitioning, but additionally provides a simpler setting to inspect the alignment of the discovered communities with the two subfields.

To do this, we construct eight pairs of subfield networks where every node has at least one of the select subfields in its list of all subfields it belongs to. Our criterion for choosing the select pairs of subfields is such that the pairs are not too overlapping, nor completely disparate both in terms of subject matter and network connectivity. We furthermore ensure the subfield communities are comparable in size.

In addition to the information reported in Table~\ref{tab:OA-field-networks}, we include the modularity of the bipartition based on metadata labels (Metadata $Q$) and the modularity of the best bipartition of a randomized realization of the same network drawn from the block null model. The significant gap between $Q$ and $Q_\text{randomized}$ shows that the community structure present in the network is well beyond the detectability limit of this model. See Appendix~\ref{appendix:randomization} for more information on this issue. 

On every subfield network, we perform exactly one bipartition followed by either the split-Louvain or the final-Louvain fine-tuning. As shown in Table~\ref{tab:OA-subfields-no-init} the final-Louvain fine-tuning outperforms its split-Louvain counterpart by a slight margin, possibly due to allowing for multiple movements for every node. Table~\ref{tab:OA-subfields-no-init} also shows that for all networks, the modularity value of the outcome partition with no fine-tuning is significantly smaller than that of the metadata partition. This shows that as a heuristic approximation to the modularity maximization problem~\cite{newman2013spectral}, the leading eigenvector spectral bipartitioning method can strongly benefit from additional optimization steps. This also reaffirms the idea of applying fine-tuning at every bipartitioning step, as opposed to a single application of a fine-tuning step after partitioning.

Unlike synthetic networks, there are generally no planted communities in real-world networks, nor the true generative process is known~\cite{peel2017ground}. Absent a ground-truth partition, metadata, typically node attributes, is often treated as such. The assumption, nevertheless, that metadata is inherently linked with the community structure cannot be readily made~\cite{peel2017ground, newman2016structure, hric2016network}.
Alongside the deep-seated reasons that metadata does not always correlate with network structure, the process of generating metadata labels can be noisy and/or biased. In the OpenAlex database, for instance, citation relationship is a feature for the classifier that produces the node labels. 
With said caveats in mind, for randomly selected networks we evaluate whether partitions found by our model and the DAG~\cite{speidel2015community} bear any resemblance to that of the metadata partitions.

For a random sample of 200 OpenAlex subfields, we construct all possible networks containing exactly two subfields. We exclude networks where one subfield contributes less than 3\% of the largest weakly connected component, totaling 13\,361 networks. The networks thus have two innate communities with varying degrees of separatedness. We then perform a spectral bipartitioning followed by the greedy split-Louvain fine-tuning step. We thereafter measure how much the fine-tuned membership vectors differ from the metadata subfield bipartition under the DAG and block-corrected modularities.

\begin{figure}[!tb]
  \centering
  \includegraphics[width=\linewidth]{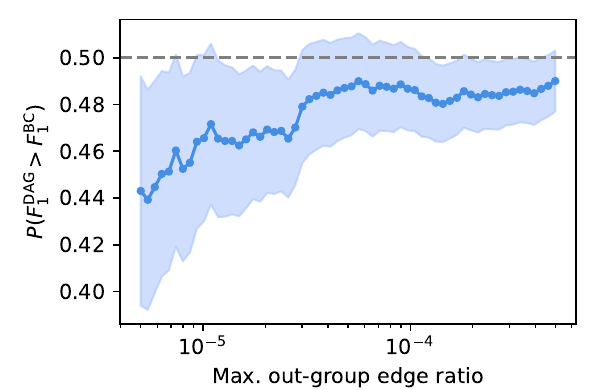}
  \caption{Probability of the DAG model producing outcomes closer to the metadata compared to the block-corrected, as measured by $F_1$ scores, decreases with the minimum threshold on separatedness of the metadata communities. The $F_1$ scores are produced by performing spectral bipartitioning, followed by split-Louvain fine-tuning (see Sec.~\ref{subsec:split-fine-tuning}).
  Note that the separatedness on the horizontal axis is plotted cumulatively.
  Shaded area shows 95\% confidence interval.}
  \label{fig:f1}
\end{figure}

For each modularity, we compute a global $F_1$ score (based on the total number of true positives, false negatives and false positives) indicating how well they recover the two original subfields, to inspect how the $F_1$ scores of the two models compare for networks with different levels of separatedness, defined as the fraction of out-group edges to all edges. To illustrate this, in Fig.~\ref{fig:f1}, we plot the probability that Block-corrected performs better than DAG for different maximum thresholds of separatedness. As the separatedness of the two subfields accentuates, the labeling advantage of block-corrected compared to DAG increases. This shows that the block-corrected modularity is more effective at recovering the partitions imposed by the omitted (unknown) subfield metadata from the network structure compared to DAG, especially in networks where innate communities imposed by subfields are more distinct. Here the choice of $F_1$ score is justified since the problem at hand is a binary classification. However, in a more general setting, a more established metric for partition comparison such as the normalized mutual information is more fitting. We additionally repeated this analysis using normalized mutual information as the measure of performance with similar results, presented in Appendix~\ref{appendix:nmi}.

\section{Conclusion and discussion}
Complex networks often include node attributes that can impact their community structure. We showed that modularity-based community detection methods can fail to reveal connectivity patterns driven by unknown attributes if known attributes and their effects on connectivity are not explicitly assimilated into the modularity framework. We introduced a new modularity measure with its associated null model, which isolates the effect of known attributes to uncover underlying structures shaped by unknown ones. This premise runs counter to most literature on community detection methods, which typically treat the attributes as ground truth or leverage them for improved results in terms of their alignment with the discovered structures.

We demonstrated the effectiveness of our proposed modularity on select temporal network models, varying in temporal connectivity patterns. We took temporal networks, specifically citation networks, as our primary application as they display an innate block structure due to the constraints imposed by the arrow of time, as well as the heavy-tail citation patterns they may exhibit. Our proposed block-corrected modularity, a variant adaptable to arbitrary temporal connectivity distributions, addresses the limitations of existing modularity-based approaches when dealing with networks characterized by the aforementioned heavy-tail edge decay.

Using the OpenAlex dataset we further showed the efficacy of our method on a range of citation networks representing select fields, and similarly on a wide array of subfield networks consisting of exactly two subfields. In the absence of an all-encompassing ground truth community structure~\cite{peel2017ground} and given the complex relationship of available metadata to structural communities~\cite{hric2014community, mangold2024quantifying}, it remains a challenge to assess and compare the quality of detected communities across methods in real-world settings. Construction of the subfield networks allowed for a more controlled experiment for metadata evaluation where we could also control for the amount of separatedness between the two subfields.

We also outlined some shortcomings of the commonly employed leading eigenvector heuristic, demonstrating in simulations how it produces suboptimal outcomes under certain conditions.
In such circumstances, methods such as greedy fine-tuning algorithms as a post-processing step can notably improve the final results. The implementation and the necessary instructions for reproducing the results of the paper are provided at Ref.~\cite{codezenodo}.

More fundamentally, community detection methods based on modularity maximization suffer from a systematic flaw. Namely, that in finite networks, a modularity maximization method reliably finds partitions with non-zero modularity in realizations of its own null model \cite{guimera2004modularity, reichardt2006networks}. In Appendix~\ref{appendix:randomization} we investigate the performance of our block-corrected modularity in the face of randomized realizations drawn from the block null model.
Bayesian inference methods for community detection, for example those based on exponential random graph models \cite{peixoto2019bayesian}, can be inherently resistant to this issue at the cost of implementation complexity. These methods can be adapted to discount the effect of known attributes on connectivity, similar to what we discussed in this paper.

Networks are often accompanied by not one, but a multitude of known attributes. The method presented here allows us to incorporate any subset of the known attributes into the null model as necessitated by the research question. 
We then treat the resulting communities as what mirrors the unknown attributes. Likewise, the resulting community structure is often affected by a combination of multiple unknown attributes. The distilling of the discovered structures into their attribute constituents lies beyond the capabilities of our model. We treat the structures we find as an all-encompassing outcome of what is likely an amalgam of unknown attributes, ones we cannot disentangle within our model definition.

\begin{acknowledgments}
We acknowledge the EuroHPC Joint Undertaking for awarding this project access to the EuroHPC supercomputer LUMI, hosted by CSC (Finland) and the LUMI consortium. 
We also acknowledge the computational resources provided by the Aalto Science-IT project. MK acknowledges grant number 349366 from the Research Council of Finland.
\end{acknowledgments}

\appendix

\section{Computing modularity values across known and unknown attributes}\label{appendix:recovering-unknown-block-structure}
Under the framework discussed in Sec.~\ref{subsec:intersecting-blocks}, the modularity value of the partition splitting the network into two groups of nodes with attributes (labels) $x = 0$ and $x = 1$ denoted by $Q_x$ is
\begin{equation}
    Q_x = \frac{1}{m} \left( \sum_{\substack{i, j \\ x_i = x_j}} A_{ij} - \sum_{\substack{i,j \\ x_i = x_j}} \Tilde{p}_{ij} \right)\,,
\end{equation}
where $\Tilde{p}_{ij}$ is the edge probability between nodes $i$ and $j$ under a null model. Modularity value of partition splitting the network into nodes with attribute $y = 0$ and $y = 1$, $Q_y$ is defined similarly as above, with $x_i$ (resp. $x_j$) replaced by $y_i$ (resp. $y_j$).

In what follows, we derive the modularity values $Q_x$ and $Q_y$ of the block-corrected modularity and the directed modularity.

Under the block null model as defined in Eq.~\eqref{eq:block-cor}, the term $k_i^\mathrm{out}/K_{x_i}^\mathrm{out}$ in $\Tilde{p}_{ij}$ becomes $2/|V|$ since there are $|V|/2$ nodes in community $x_i$, the known attribute. Thus $\Tilde{p}_{ij}$ transforms into $4/|V|^2 L_{x_ix_j}$ where $L_{x_ix_j}$ is the number of edges between nodes with label $x_i$ and nodes with label $x_j$.

We assume edge densities in each community of $x = 0$ and $x = 1$ are equal, as well as between the two, that is, $L_{00} = L_{11}$ and $L_{01} = L_{10}$. 
Node $i$ can be connected to a node whose label for $y$ could be the same with probability $p_1^y$, or different with probability $p_0^y$. There are $|V|/4 - 1$ possible nodes tagged with the same $y$ as node $i$ and $|V|/4$  with a different value for $y$. Since there are $|V|/2$ possible nodes with the same value for $x$ as node $i$ we get
\begin{align*}
    L_{00} &= L_{11} \\
    &= \frac{|V|}{2} p_1^x((\frac{|V|}{4} - 1)p_1^y + \frac{|V|}{4} p_0^y) \\
    &\approx \frac{|V|^2}{8} p_1^x(p_1^y + p_0^y)\,.
\end{align*}

Following a similar logic, $\sum_{\substack{i, j \\ x_i = x_j}} A_{ij}$ is approximately equal to $ |V|^2 p_1^x(p_1^y + p_0^y)/4$. We substitute this and $\Tilde{p}_{ij} = p_1^x(p_1^y + p_0^y)/2$ into $Q_x$ to get
\begin{align*}
   m &Q^\mathrm{bc}_x = \frac{|V|^2}{4} p_1^x(p_1^y + p_0^y) -  \sum_{\substack{i,j \\ x_i = x_j}} \frac{p_1^x(p_1^y + p_0^y)}{2} \\
    &= \frac{|V|^2}{4} p_1^x(p_1^y + p_0^y) - p_1^x(p_1^y + p_0^y)\frac{|V|}{2} (\frac{|V|}{2} - 1) \approx 0\,.
\end{align*}
This approximation holds for $|V| \gg 0$.

Using our proposed modularity, the partition dividing the network with respect to the known attribute $x$ thus provides no improvement over a single partition covering the entire network in terms of modularity maximization, making such bipartitioning undesirable.

Computing $Q^\mathrm{bc}_y$, the term $\sum_{i, j; y_i = y_j} A_{ij}$ is the same as its $Q^\mathrm{bc}_x$ equivalent, with $x$ and $y$ replaced:
\begin{equation}\label{eq:Q_y_long}
    Q^\mathrm{bc}_y = \frac{1}{m} \left( \frac{|V|^2}{4} p_1^y(p_1^x + p_0^x)  - \sum_{\substack{i,j \\ y_i = y_j}} \Tilde{p}_{ij} \right)\,.
\end{equation}

Elements $L_{01} = L_{10}$ of matrix $\mathbf{L}$ are equal to $|V|/2(|V| p_0^x/4 + p_1^y + |V| p_0^x p_0^y/4) = |V|^2 p_0^x(p_0^y + p_1^y)/8$ since there are $|V|/2$ nodes in each community with respect to $x$ and for every node the destination node is in the same community with respect to $y$ with probability $p_1^y$ and probability $p_0^y$ of being in different communities with respect to $y$.

We can write out $\sum_{i,j; y_i = y_j} \Tilde{p}_{ij}$ as
\begin{equation}\label{eq:eq-sums}
    \sum_{\substack{i,j \\ y_i = y_j}} \Tilde{p}_{ij} = 
    \sum_{i=1}^{|V|} \sum_{j; y_j = y_i} \Tilde{p}_{ij}\,.
\end{equation}

The sum $\sum_{j; y_j = y_i} \Tilde{p}_{ij}$ is divided into cases where $x_i$ and $x_j$ are equal or otherwise, that is,
\begin{align*}
    (\frac{|V|}{4} - 1)(\frac{4}{|V|^2} L_{00}) + \frac{|V|}{4}\frac{4}{|V|^2} L_{01} \approx \frac{1}{|V|} (L_{00} + L_{01})\,.
\end{align*}

Plugging this in Eq.~\ref{eq:eq-sums} we get
\begin{equation}
    \sum_{\substack{i,j \\ y_i = y_j}} \Tilde{p}_{ij} =
    \sum_{i=1}^{|V|} \frac{1}{|V|} (L_{00} + L_{01}) = L_{00} + L_{01}\,.
\end{equation}
We substitute the right-hand side of Eq.~\eqref{eq:eq-sums} in Eq.~\eqref{eq:Q_y} and plug in $2(L_{00} + L_{01}) = |V|^2(p_1^x + p_0^x)(p_1^y + p_0^y)/4$ for $m$ to arrive at
\begin{equation}\label{eq:Q_y_final}
    Q^\mathrm{bc}_y = \frac{1}{2}\frac{p_1^y - p_0^y}{p_1^y + p_0^y}\,.
\end{equation}
Modularity value $Q^\mathrm{bc}_y$ in Eq.~\eqref{eq:Q_y_final} increases as the probability of two nodes having the same values for label $y$ approaches one ($p_1^y \rightarrow 1$) and the probability of two nodes having different values for $y$ approaches 0 ($p_0^y \rightarrow 0$). In other words, the more separated the two communities of $y=0$ and $y=1$, the larger the value for $Q^\mathrm{bc}_y$, thus rendering such a division a more desirable one.

The probability of an edge in the directed model is defined as $\Tilde{p}_{ij} = \frac{k_i^{out}k_j^{in}}{m}$, solely dependent on in- and out-degrees of the two nodes of an edge. This indicates modularity values $Q^\mathrm{dir}_x$ and $Q^\mathrm{dir}_y$ are calculated symmetrically, i.e., $Q^\mathrm{dir}_y$ can be derived through replacing $p^x$ with $p^y$ in $Q^\mathrm{dir}_x$. Deriving $Q^\mathrm{dir}_x$ alone should thus suffice. The average in-/out-degree would be $|v|/4(p^x_0 p^y_0 + p^x_0 p^y_1 + p^x_1 p^y_0 + p^x_1 p^y_1)$ which we refer to as $|V|^2 p^{xy}/4$ for brevity, and the number of edges $m$ is computed similarly as before, equal to $|V|^2 (p_1^x + p_0^x)(p_1^y + p_0^y)/4 = |V|^2 p^{xy}/4$. Modularity value $Q^\mathrm{dir}_x$ is thus
\begin{align*}
    &Q^\mathrm{dir}_x = \frac{1}{m} \frac{|V|^2}{4} p_1^x(p_1^y + p_0^y) - \frac{1}{m^2} \sum_{\substack{i,j \\ x_i = x_j}} \frac{|v|^2}{16} (p^{xy})^2 \\
    &=  \frac{1}{m} \frac{|V|^2}{4} p_1^x(p_1^y + p_0^y) - \frac{1}{m}\sum_{\substack{i,j \\ x_i = x_j}} \frac{p^{xy}}{4} \\
    &\approx \frac{1}{m} \frac{|V|^2}{4} p_1^x(p_1^y + p_0^y) - \frac{1}{m}\frac{|V|^2}{2}\frac{p^{xy}}{4} \\
    &= \frac{1}{2}\frac{p_1^x - p_0^x}{p_1^x + p_0^x}\,.
\end{align*}

Modularity value $Q^\mathrm{dir}_y$ therefore is
\[
Q^\mathrm{dir}_y = \frac{1}{2}\frac{p_1^y - p_0^y}{p_1^y + p_0^y}\,.
\]

It might be argued that a bipartitioning based on known attribute $x$ by the directed modularity might still help in recovering the unknown attribute, if the modularity is then allowed to subsequently bipartition the two discovered communities into four communities, based on values of both $x$ and $y$. While this four-community solution certainly works for the example presented in Fig.~\ref{fig:intersecting_schematic}, it is not generally the case. Starting from Eq.~\eqref{eq:modularity_func}, calculating the modularity of the four-community solution with directed and block-corrected modularity similar to the derivations above, we get
\begin{equation}
    Q^\mathrm{dir}_{xy} = \frac{p_1^y p_1^x}{(p_1^y + p_0^y)(p_1^x + p_0^x)} - \frac{1}{4}\,,
\end{equation}
and
\begin{equation}
    Q^\mathrm{bc}_{xy} = \frac{1}{2}\frac{p_1^x (p_1^y - p_0^y)}{(p_1^y + p_0^y)(p_1^x + p_0^x)}\,.
\end{equation}

When the inequalities
\begin{equation}
    Q^\mathrm{bc}_{y} > Q^\mathrm{bc}_{x} \ \text{and}\ 
    Q^\mathrm{bc}_{y} > Q^\mathrm{bc}_{xy}
\end{equation}
are held, the modularity maximization method using the proposed block-corrected modularity will successfully recover hidden attribute $y$, but does not continue to needlessly split the found communities based on $(x, y)$. Since $Q^\mathrm{bc}_{x} \approx 0$, the first part of the inequality simplifies into $p^y_1 > p^y_0$, i.e., that the hidden attribute must be assortative. The second inequality translates to
\begin{equation*}
    \frac{1}{2}\frac{p_1^y - p_0^y}{p_1^y + p_0^y} > \frac{1}{2}\frac{p_1^x (p_1^y - p_0^y)}{(p_1^y + p_0^y)(p_1^x + p_0^x)}, 
\end{equation*}
from which we obtain $p_0^x > 0$.

Similarly, when inequalities
\begin{equation}
    Q^\mathrm{dir}_{x} > Q^\mathrm{dir}_{y} \ \text{and}\ 
    Q^\mathrm{dir}_{x} > Q^\mathrm{dir}_{xy}
\end{equation}
are held, the modularity maximization method using the directed modularity will not split the initial (erroneous) bipartitioning based on the known attribute into four communities based on $(x, y)$ values. From the first inequality, we have
\begin{equation*}
    \frac{1}{2}\frac{p_1^x - p_0^x}{p_1^x + p_0^x} > \frac{1}{2}\frac{p_1^y - p_0^y}{p_1^y + p_0^y}, 
\end{equation*}
which simplifies to $p_1^x p_0^y > p_0^x p_1^y$.
As we assumed $p^y_1 > p^y_0$ before, the first inequality is held if known attribute $x$ is more strongly assortative than unknown attribute $y$. From the second inequality, we obtain
\begin{equation*}
    \frac{1}{2}\frac{p_1^x - p_0^x}{p_1^x + p_0^x} > \frac{p_1^y p_1^x}{(p_1^y + p_0^y)(p_1^x + p_0^x)} - \frac{1}{4}, 
\end{equation*}
which leads to $p_1^x (3 p_0^y - p_1^y) > p_0^x (p_1^y + p_0^y)$. 
Since the left-hand side is positive, 
this implies that $3 p_0^y - p_1^y > 0$ should hold and
\begin{equation}
    \frac{p_1^x}{p_0^x} > \frac{p_1^y + p_0^y}{3p_0^y - p_1^y}\,.
\end{equation}

For example, with $p_1^x = 0.9$, $p_0^x = 0.05$, $p_1^y = 0.35$ and $p_0^y = 0.25$, we get $Q^\mathrm{dir}_{x} = 0.45$, $Q^\mathrm{bc}_{x} \approx 0$, $Q^\mathrm{dir}_{y} = Q^\mathrm{bc}_{y} = 0.084$, $Q^\mathrm{dir}_{xy} = 0.30$ and $Q^\mathrm{bc}_{xy} = 0.078$ which satisfies both inequalities above. In this scenario, directed modularity splits the network into two communities based on the known attribute $x$ but does not proceed to split it again into four communities based on $(x, y)$, while block-corrected modularity correctly identifies the unknown attribute $y$ and does not further divide the network.

\section{Power iteration tolerance scaling with network parameters}\label{appendix:tol}
Spectral clustering through calculating exact values for the leading eigenvectors of the dense modularity matrix is a computationally expensive task. Employing the power iteration method for finding the leading eigenvector can not only help utilize sparse matrix operations but also provide a vehicle for vectorisation, which in turn significantly improves computation time. This is described in detail in Sec.~\ref{subsec:vectorised-px}.

Different implementations of the power iteration algorithm rely on different convergence criteria. Often, the algorithm halts when the estimated eigenvalue remains within a certain threshold of that of the previous iteration. This threshold is here denoted as tolerance. A smaller value for tolerance is a stricter condition on convergence for the eigenvector.

\begin{figure}[!tb]
  \centering
  \includegraphics[width=\linewidth]{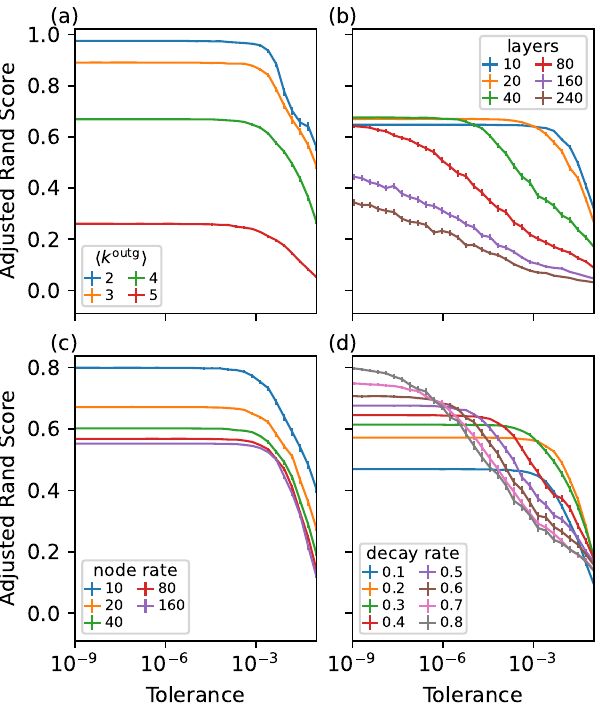}%
  \caption{Quality of bipartitioning proposed by the spectral method for random networks containing two planted communities as a function of the power iteration tolerance. Average in-group degree (a) and number of nodes in each time layer (c) do not change the required tolerance threshold, while an increase in the number of time layers (b), or higher prevalence of longer edges (d) necessitates smaller values for tolerance. Unless stated in the plot, the generated networks have 20 nodes per time layer, 20 time layers (40 for (d)), in-group degree 8, out-group degree 4, and an edge decay rate of 0.5.}
  \label{fig:tol}
\end{figure}

The choice for the tolerance value affects both runtime and the precision of the converged eigenvector. It thus raises the question of what value we must choose for tolerance. We, moreover, observe in our study of empirical networks in Sec.~\ref{sec:openalex} that fluctuations in the number of communities occur across multiple runs of the community detection process, indicating variance among the power iteration outcomes across runs. Motivated by these considerations, we attempt to study the interplay between tolerance and the structural properties of the network and the impact it may have on the community detection outcomes.

Our goal is to explore the relationship between various network parameters and the required tolerance value to obtain a certain level of accuracy. The ensemble of networks is drawn from the temporal network model with two planted communities and exponentially distributed edge lengths, details of which are discussed in Sec.~\ref{sec:temporal-block-structure-ex}.

Illustrated in Fig.~\ref{fig:tol}, it is evident that the tolerance has a clear relationship with the number of temporal layers. As networks become temporally longer (Fig.\ref{fig:tol}(b)) or as longer edges become more prevalent (Fig.\ref{fig:tol}(d)), the required tolerance to maintain the accuracy level tightens. For larger and longer networks, this can pose an obstacle as the accuracy of computing the difference in eigenvalues of consecutive iterations is bounded by floating-point errors. In practical terms and for the case of real-world citation networks, we observe that tolerance values smaller than $10^{-13}$ fail to reliably converge within a reasonable number of iterations.

\section{Temporal edge length distribution in OpenAlex citation 
 networks}\label{appendix:delta-distribution}

 \begin{figure}[!htb]
  \centering
  \includegraphics[width=\linewidth]{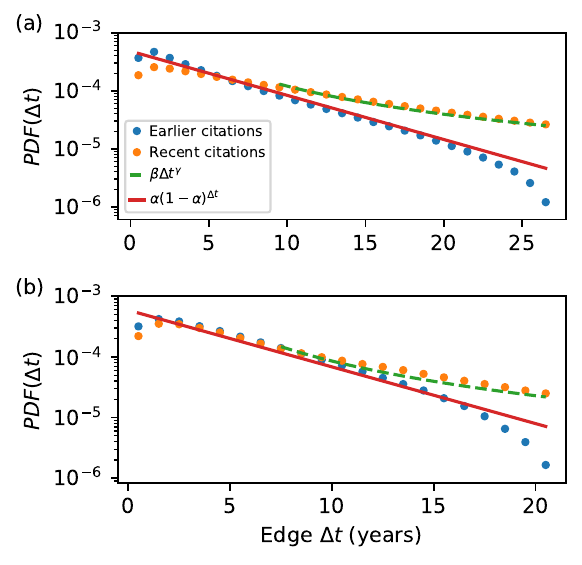}%
  \caption{Distribution of temporal edge lengths in years for citation networks of (a) Physics and Astronomy and (b) Computer Science fields in the OpenAlex database. Earlier citations include references made by papers published before the year 2000. Recent citations include references spanning 9 years 2015--2024. An exponential is fitted to the earlier citation distributions and a power law to the tail of the recent citation distributions. Fitted curve parameters are (a) $\beta = 63.42 $, $\gamma = -1.6$, $\alpha = 4.8 \times 10^{-4}$ and (b) $\beta = 5.0 \times 10^3 $, $\gamma = -1.9$, $\alpha = 5.9 \times 10^{-4}$. The probability distribution of recent citations as a function of $\Delta t$ has a heavier tail compared to that of the earlier citations.}
  \label{fig:OA-deltas}
\end{figure}

As shown in Figure~\ref{fig:OA-deltas}, the distributions of temporal edge lengths in both networks seem to be better fitted by an exponential distribution for earlier publications, whereas for more recent publications, published in the past roughly 10 years (2015-2024), they are more suitably fitted by a power law decay. The results indicate that connectivity patterns induced by time in citation networks have undergone significant change, signaling that, similar to the examples studied in Sec.~\ref{sec:temporal-block-structure-ex}, the applied community detection method should be able to (1) take into account this temporal connectivity pattern, and (2) be able to flexibly handle this changing connectivity pattern for it to be effective.

\section{Subfield networks evaluation using Normalized Mutual Information}\label{appendix:nmi}

Figure~\ref{fig:f1} reports the agreement between the automatically detected communities and the subfield labels through the $F_{1}$ score. The choice of $F_{1}$ score is adequate as long as we treat the task as binary classification, i.e., if we know \emph{a priori} that every article belongs to one of exactly two known subfields and we care equally about precision and recall. Once we step outside this binary framework, however, the use of a classification metric becomes problematic because (i) the community labels obtained from an unsupervised algorithm are arbitrary up to permutation, and (ii) the number of inferred communities need not equal the number of ground-truth classes. Normalized mutual information (NMI) sidesteps both issues by quantifying information overlap between two partitions independently of label identities and cardinalities, making it the standard choice for partition-level evaluation in community detection research.

Despite its widespread use, normalized mutual information is not, however, without shortcomings. Chiefly, it exhibits bias in finite samples, that is, partitions with many small communities can achieve deceptively high normalized mutual information against a reference partition with heterogeneous community sizes, and it effectively ignores part of the information encoded in the contingency table~\cite{jerdee2023normalized}.

\begin{figure}
    \centering
    \includegraphics[width=\linewidth]{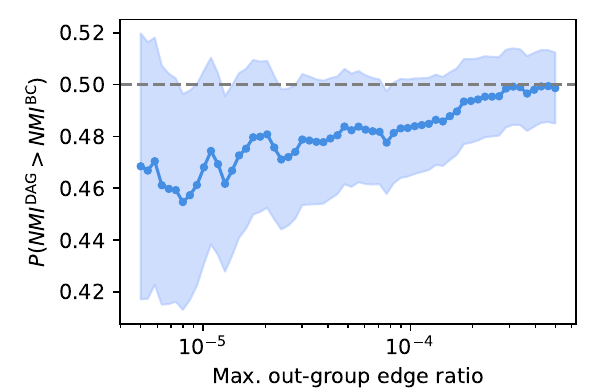}
    \caption{Probability of the DAG model producing outcomes closer to the metadata compared to the block-corrected, as measured by normalized mutual information (NMI), decreases with the minimum threshold on separatedness of the metadata communities. This confirms the similar observation in Fig.~\ref{fig:f1} based on $F_1$ scores. The NMIs are produced by performing spectral bipartitioning, followed by Split fine-tuning (see Sec.~\ref{subsec:split-fine-tuning}). Note that the separatedness on the horizontal axis is plotted cumulatively. Shaded area shows 95\% confidence interval.}
    \label{fig:nmi}
\end{figure}

Figure~\ref{fig:nmi} provides a reproduction of Fig.~\ref{fig:f1}, using normalized mutual information instead of $F_1$ scores for comparing labels inferred from the data. This confirms that our presented method proposes communities that are closer to the metadata communities compared to the DAG model, as measured by larger normalized mutual information between metadata and detected communities.

\section{Detectability limit and overfitting}\label{appendix:randomization}

Any network drawn from the block null model preserves layer structure and layer connectivity of the original network, as well as the in- and out-degree distributions (on expectation), but without any community structure other than what is induced by the layer structure. A randomized realization of a null model is a network drawn from the null model with its edges shuffled and the same properties preserved. Ideally, a modularity-based community detection method should detect no significant community structure on realizations of its own corresponding null model. In actuality, nevertheless, modularity maximization methods often find communities with non-negligible modularity values where none exist~\cite{guimera2004modularity, reichardt2006networks}.

\begin{figure}
    \centering
    \includegraphics[width=\linewidth]{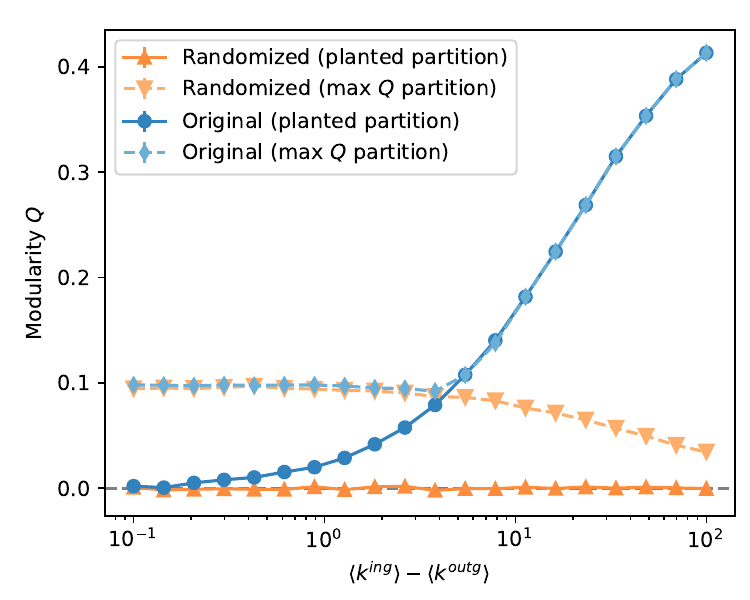}
    \caption{Modularity value of the planted partition (dark blue) and the maximum modularity partition (light blue) under the block-corrected modularity over an ensemble of random networks with varying average in-group degrees. All networks are realizations of the exponential model (Eq.~\eqref{eq:exponential-model}), with $t_\mathrm{max} = 12$ time layers, $N = 100$ nodes per layer, $b = 2$ planted communities, and an edge decay rate of $d = 0.5$.
     Error bars show 95\% confidence interval. When observing the modularity values of the same partitions in the null model realizations (dark and light orange) we see that below a threshold of distinctness, the true partition is not detectable.}
    \label{fig:randomized}
\end{figure}

For an array of random networks drawn from the exponential model described by Eq.~\eqref{eq:exponential-model}, Fig.~\ref{fig:randomized} inspects how the block-corrected modularity would treat the randomized versions of the networks and whether it detects the true partitions.
For both the original network and its randomized counterpart, we compute the modularity of the true planted partition and the partition resulting from maximizing the modularity function.

As shown in Fig.~\ref{fig:randomized}, below a certain threshold of distinctiveness between communities, the modularity values of the maximum modularity partition on both original and the randomized networks are on par, an indication that the true partition would not be detectable and that the maximum modularity partition is explained by the tendency of the modularity maximization method to overfit. However, as the difference between the in-group and out-group degrees becomes greater the maximum modularity partition converges to that of the true planted partition, indicating that the true partition becomes detectable.

\section{Comparison with Ordered Stochastic Block Model}\label{appendix:osbm}

Thus far we compared the performance of our proposed block-corrected modularity to other modularity based approaches. There are, however, other methods not necessarily based on modularity, that take an inherent ordering present in the network into account. A primary example is Ordered Stochastic Block Model (OSBM) and its degree-corrected variant (DC-OSBM)~\cite{peixoto2022ordered}. A major difference between the approach presented here and that of Ref.~\cite{peixoto2022ordered}, however, is that the block-corrected modularity takes any inherent ordering imposed on nodes as an \emph{a priori} attribute, readily available as an input to the model, while the latter attempts to infer the ordering, as an additional layer on top of the inferred communities.

\begin{figure}
    \centering
    \includegraphics[width=\linewidth]{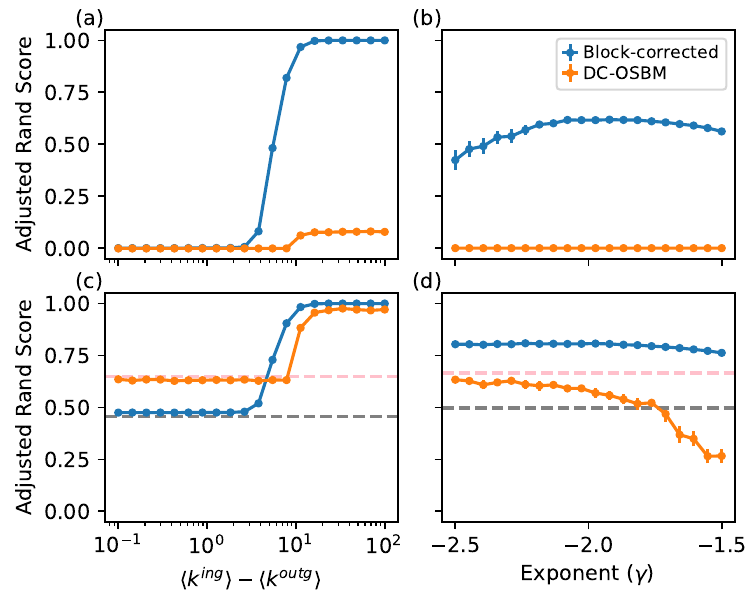}
    \caption{Adjusted Rand scores between discovered communities and the true partitions.
    Realizations of the exponential model (Eq.~\eqref{eq:exponential-model}) are networks with $t_\mathrm{max} = 12$ time layers, $N = 100$ nodes per layer, $b = 2$ planted communities, and an edge decay rate of $d = 0.5$ (panels (a) and (c)). Realizations of the power law model (Eq.~\eqref{eq:power-law-model}) are networks with $t_\mathrm{max} = 200$ time layers, $N = 100$ nodes per layer, $b = 2$ planted communities, in-group degree 8, and out-group degree 4 (panels (b) and (d)). For panels (c) and (d) the outcome communities detected by block-corrected modularity are split based on the time layers for parity with DC-OSBM. The gray dashed line shows the expected Adjusted Rand Score for correctly inferring all node-times while randomly assigning the unknown attribute. The pink dashed line indicates expected Adjusted Rand Score corresponding to correctly inferring only the $t_\mathrm{max}$ time layers. In panels (a) and (b) the comparison is done on the raw output communities. Error bars show 95\% confidence intervals.}
    \label{fig:osbm}
\end{figure}

Despite this crucial difference we seek to investigate the models' performances on an ensemble of synthetic networks drawn from the exponential generative model (Eq.~\eqref{eq:exponential-model}) across varying average in-group degrees, and the power law model (Eq.~\eqref{eq:power-law-model}) across exponents. We show the comparison of performance of the block-corrected model and DC-OSBM in Fig.~\ref{fig:osbm}(a,b). The direct comparison of the resulting partitions based on Adjusted Rand Score with the planted partition shows that the block-corrected modularity is more successful at directly inferring the hidden attribute.

Considering that the OSBM and DC-OSBM attempt to infer the time layers as well as the planted partition, on our synthetic network models they ideally infer $t_\mathrm{max} \times b$ distinct communities. 
To make a comparison that does not disadvantage the methods that split communities in time, we adapt the ground-truth and the discovered communities of the block-corrected method by dividing them into their constituent time layers, forming $t_\mathrm{max} \times b$ communities.
This should somewhat balance the fact that OSBM-based methods are at a disadvantage due to the method not being designed for this particular problem. However, it should be noted that in real-world use cases, we would want to infer communities that are not split in time.
The results are illustrated in Fig.~\ref{fig:osbm}(c,d).

In the exponential model based on Eq.~\eqref{eq:exponential-model}, the block-corrected modularity consistently outperforms the DC-OSBM when comparing the raw partitions (Fig.~\ref{fig:osbm}(a) and (b)). In the case where ground-truth communities are split into $t_\mathrm{max} \times b$ communities, shown in Fig.~\ref{fig:osbm}(c), for networks with only a small difference between in- and out-group degrees the block-corrected modularity fails to distinguish anything other than the time layers, as they coincide with the dashed gray line indicating correctly inferring the time layers and random assignment for the ground-truth communities. This can be ascribed to the detectability limit described in Appendix~\ref{appendix:randomization}.
The DC-OSBM, however, levels at the analytical expected adjusted Rand score for inferring the time layers only (dashed pink line), forming a total of $t_\mathrm{max}$ communities. As the planted communities become more distinct, the classification performance of both models rise toward 1.0, while the block-corrected achieves slightly better outcomes.

For the case of power law edge length model based on Eq.~\eqref{eq:power-law-model}, shown in Fig.~\ref{fig:osbm}(d), the block-corrected modularity surpasses the DC-OSBM across all values of the exponents of the power law edge length distributions. Notably, the DC-OSBM fails to correctly infer the time layers for particularly heavy-tail edge length distributions, falling below the dashed gray line.

\bibliography{references}
\end{document}